\documentclass[useAMS,usegraphicx]{mn2e}
\usepackage{times}
\usepackage{rotate}
\usepackage{amssymb}
\usepackage{multirow}
\title[The role of AGN in the colour transformation of galaxies at
  redshifts $z \approx 1$]
{The role of AGN in the colour transformation of galaxies at redshifts $z
  \approx 1$}   
\author[Georgakakis et al. ] {A. Georgakakis$^{1}$\thanks{Marie Curie
    fellow},  K. Nandra$^{1}$, R. Yan$^{2}$,  S. P. Willner$^3$, J. M. Lotz$^{4,5}$, C. M. Pierce$^{4}$,\\\\
    {\rm \LARGE  M. C. Cooper$^{2,6}$, E. S. Laird$^{1}$, D. C. Koo$^7$,
      P. Barmby$^3$, J. A. Newman$^8$, J. R. Primack$^{9}$}\\\\
    {\rm \LARGE A. L. Coil$^{6}$}\\\\
  $^1$Astrophysics Group, Blackett Laboratory, Imperial College, Prince
  Consort Rd , London SW7 2BZ, UK\\
  $^2$Department of Astronomy, University of California, Berkeley, CA 94720\\ 
  $^3$Harvard-Smithsonian Center for Astrophysics, 60 Garden Street,
  MailStop 65, Cambridge, MA02138, USA\\
  $^4$Department of Physics, University of California, Santa Cruz,
  1156 High Street, CA 95064,  USA\\ 
  $^5$National Optical Astronomical Observatories, 950 N. Cherry
  Avenue, Tucson, AZ 85719, USA \\
  $^6$Steward Observatory, University of Arizona, 933 N. Cherry Ave.,
  Tucson, AZ 85721-0065, USA\\
  $^7$UCO/Lick Observatory and Department of Astronomy \& Astrophysics, University of California, Santa Cruz,  CA
  95064, USA\\  
  $^8$Lawrence Berkeley National Laboratory, Berkeley, CA 94720, USA\\
   $^9$Department of Physics, University of California, Santa Cruz,
  1156 High St., Santa Cruz, CA 95064, USA\\
}
\begin{document}
\maketitle  

\begin{abstract} 

We  explore the  role of  AGN in  establishing and/or  maintaining the
bimodal   colour   distribution  of   galaxies   by  quenching   their
star-formation and  hence, causing their  transition from the  blue to
the  red cloud.   Important tests  for this  scenario include  (i) the
X-ray properties  of galaxies in  the transition zone between  the two
clouds and (ii) the incidence of AGN in post-starbursts, i.e.  systems
observed   shortly  after   ($<1$\,Gyr)  the   termination   of  their
star-formation.   We  perform  these  tests  by  combining  deep  {\it
Chandra} observations with multiwavelength data from the AEGIS survey.
Stacking the X-ray photons  at the positions of galaxies ($0.4<z<0.9$)
not individually  detected at X-ray wavelengths  suggests a population
of obscured  AGN among sources in  the transition zone and  in the red
cloud.   Their mean X-ray  and mid-IR  properties are  consistent with
moderately obscured low-luminosity AGN, Compton thick sources or a mix
of both.  Morphologies  show that major mergers are  unlikely to drive
the evolution  of this  population but minor  interactions may  play a
role.  The  incidence of  obscured AGN in  the red cloud  (both direct
detections and  stacking results) suggests that  BH accretion outlives
the termination of the star-formation.   This is also supported by our
finding that  post-starburst galaxies at  $z \approx 0.8$ and  AGN are
associated,  in agreement  with recent  results at  low-$z$.   A large
fraction of  post-starbursts and red cloud galaxies  show evidence for
at least moderate levels of AGN obscuration.  This implies that if AGN
outflows  cause  the  colour  transformation of  galaxies,  then  some
nuclear gas and  dust clouds either remain unaffected  or relax to the
central galaxy regions after quenching their star-formation.
\end{abstract}

\section{Introduction}\label{sec_intro}
A major recent development in extragalactic astronomy is the discovery
that  most  of   the  spheroids  in  the  local   Universe  contain  a
super-massive black hole  (BH; e.g. Magorrian et al.   1998). The masses
of  these  monsters  are tightly  correlated  to  the  stellar  velocity
dispersion of  the host  galaxy bulges (e.g.   Ferrarese et  al.  2000;
Gebhardt et al. 2000), suggesting  that the formation and evolution of
spheroids and the  build-up of the super-massive BHs  at their centres
are interconnected.  This interplay may hold the key for understanding
some of the observed trends in galaxy evolution.

One  of  the  fundamental  properties  of galaxies  is  their  bimodal
distribution  in rest-frame  colour.  This bimodality  is observed  to
beyond redshift $z\approx1$ and is believed to hold important clues on
galaxy assembly (e.g. Strateva et al.  2001; Baldry et al.  2004; Bell
et al.   2004; Weiner et al.   2005; Cirasuolo et  al. 2007).  Evolved
spheroidal   galaxies  are   found   in  the   ``red-cloud''  of   the
colour-magnitude diagram (CMD), star-forming systems define the ``blue
cloud'',  while the  region  between these  overdensities is  sparsely
populated.   This bimodal distribution  is likely  to be  the combined
result of evolution effects  and external factors such as interactions
and mergers (e.g.  Bell et al.  2004, 2006a, 2006b; Blanton 2006).  In
the  simplest interpretation,  the CMD  is consistent  with  a picture
where  the galaxy  star-formation is  truncated, by  a  mechanism that
remains  to  be identified,  leading  to  the  ageing of  the  stellar
population and  the rapid transition from  the blue to  the red cloud.
AGN are  proposed to play a  major part in this  process by regulating
the star-formation  in galaxies causing their transition  in colour or
maintaining them in the red cloud.  

Modelling  work indeed  suggests that  the energy  released by  AGN is
sufficient to  either heat up  or blow away  the cold gas  of galaxies
(e.g.   Silk  \& Rees  1998;  Fabian  1999;  King 2003),  irreversibly
altering their evolution.   Two main routes are proposed  by which AGN
feedback affects the  host galaxy.  The first one  operates during the
luminous  and  high  accretion  rate  stage of  BH  growth,  which  is
suggested to occur during major  mergers (Hopkins et al.  2005, 2006a,
b).  These  catastrophic events  also produce nuclear  starbursts that
obscure the central engine for  most of its active lifetime.  When the
AGN becomes sufficiently luminous it drives outflows, which sweep away
the  nuclear  gas  and  dust  clouds, thereby  quenching  the  nuclear
star-formation.  Following this stage, the AGN also declines as the BH
runs out of accreting material and eventually switches off.

In addition  to the  cold gas accretion  mode above,  simulations also
include a second AGN feedback recipe, which is invoked to suppress the
accretion  of hot gas  onto the  central galaxy  of large  dark matter
haloes (e.g.  Croton  et al.  2006; Okamoto et  al.  2007; Cattaneo et
al. 2007).  This  mode operates in systems where  the supermassive BHs
have  already been  built  through mergers  and  are in  place at  the
central  galaxy  regions.  In  this  picture,  as  galaxies enter  the
massive  dark matter  haloes of  groups  or clusters,  their cold  gas
supply is cut off leading  to passive evolution and their migration to
the red cloud (Dekel \&  Birnboim 2006; Croton et al. 2006).  However,
cooling flows in these dense environments could produce a reservoir of
cold gas in the central galaxy regions, the raw material for starburst
and  QSO activity.   Low-level BH  accretion is  therefore  invoked to
counterbalance cooling flows by producing low-luminosity AGN that heat
the bulk  of the  cooling gas and  thereby suppressing  any subsequent
star-formation in these galaxies.

In  the two  AGN feedback  prescriptions above,  the  outflow scenario
during  the luminous  and  high  accretion rate  phase  of BH  growth,
assigns  AGN  a  central  role  in  establishing  the  bimodal  colour
distribution of galaxies by directly causing their transition from the
blue to the red cloud.  The  distribution of X-ray selected AGN in the
CMD  appears to  be broadly  consistent  with this  picture.  AGN  are
preferentially associated with  galaxies in the red cloud,  at the red
limit of the  blue cloud and in the transition  zone in between. These
associations   are  suggestive   of  AGN-driven   truncation   of  the
star-formation (Nandra et al.  2007).  However, more detailed study of
the  properties of  AGN reveal  a number  of inconsistencies  with the
merger scenario.  Obscured AGN are predominantly hosted by galaxies in
the red cloud (Nandra et  al.  2007; Rovilos \& Georgantopoulos 2007),
with prominent  bulges and  little morphological evidence  for ongoing
major mergers  (Grogin et  al.  2003; Pierce  et al.  2007).   The red
optical colours of these systems  are most likely because of old stars
and not  dusty star-formation.  Although there are  examples of deeply
buried AGN  in star-forming  galaxies (e.g. Genzel  et al.   1998; Cid
Fernandes et al.  2001; Franceschini  et al.  2003; Georgakakis et al.
2004; Alexander  et al.  2005),  the frequency of these  systems among
the obscured  X-ray population at  $z\approx1$ is not yet  clear.  For
example, Rovilos  et al.  (2007) argue that  a non-negligible fraction
of X-ray  selected AGN at $z  \approx 1$ have  $\mu$Jy radio continuum
emission (1.4\,GHz) that is consistent with star-formation in the host
galaxy.  These  authors however,  do not find  a strong  trend between
X-ray obscuration and the incidence of faint radio emission, likely to
be associated with starbursts.

A plausible interpretation of the evidence above is that current X-ray
surveys are  biased against the early  phase of BH  evolution. At this
initial  stage of  their formation,  AGN  may be  deeply buried  under
star-forming clouds and/or low luminosity because the BH mass is still
small, although growing rapidly.  If there is such a population of
obscured  and/or intrinsically  faint  young AGN  below the  detection
threshold of  current surveys, it may  be possible to  find them using
stacking  analysis.  Alternatively,  AGN  may not  be responsible  for
quenching the star-formation in  galaxies and producing the bimodality
of the  CMD. A key test  of this scenario  is the incidence of  AGN in
post-starburst  galaxies,  i.e.  systems  observed  shortly after  the
termination  of  their   star-formation  ($<1$\,Gyr).   In  the  local
Universe  ($z  \lesssim  0.1$),  a  number  of  studies  point  to  an
association between  {\it optically} selected  AGN and post-starbursts
(e.g.  Kauffmann  et al. 2004;  Goto 2006; Yan  et al.  2006).   At $z
\approx  1$, close  to the  peak of  the AGN  density in  the Universe
(e.g. Barger et  al. 2005; Hasinger et al. 2005),  there is still very
limited  information  on  the  link  between X-ray  selected  AGN  and
post-starbursts.

We  address  the  issues  above  using data  from  the  All-wavelength
Extended Groth strip International Survey (AEGIS; Davis et al.  2007).
X-ray  stacking  is  employed  to  search  for  deeply  buried  and/or
low-luminosity AGN  among optically selected galaxies  in the redshift
interval  $0.4<z<0.9$   to  explore  the  role  of   BH  accretion  in
establishing  the bimodal  colour distribution  of these  systems.  We
also study the X-ray  properties of post-starbursts as identified from
Keck spectra at $z\approx 0.8$  to investigate the link between recent
star-formation events and active BHs.  We  adopt $\rm H_{0} = 70 \, km
\,   s^{-1}  \,   Mpc^{-1}$,   $\rm  \Omega_{M}   =   0.3$  and   $\rm
\Omega_{\Lambda} = 0.7$.

\begin{figure}
\begin{center}
 \rotatebox{0}{\includegraphics[height=0.9\columnwidth]{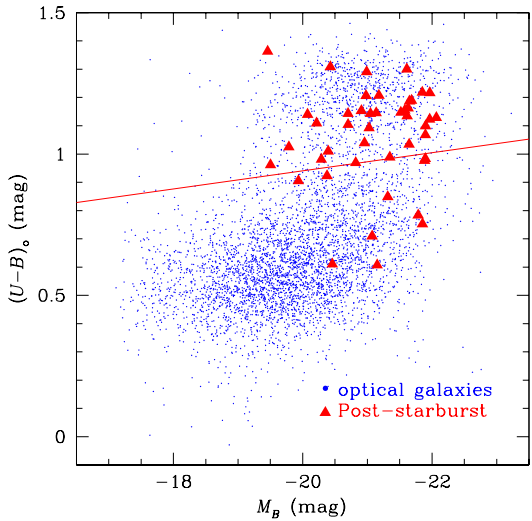}}
\end{center}
\caption{ Rest-frame $U -  B$ colour against B-band absolute magnitude
for DEEP2  galaxies (small  blue dots) in  the range  $0.4 < z  < 0.9$
estimated by Willmer et  al. (2006). Post-starburst galaxies are shown
with the (red)  triangles.  The continuous line is  defined by Willmer
et   al.  (2006)   to  separate   the  red   from  the   blue  clouds.
}\label{fig_cmd}
\end{figure}

\begin{figure*}
\begin{center}
 \rotatebox{0}{\includegraphics[height=0.7\columnwidth]{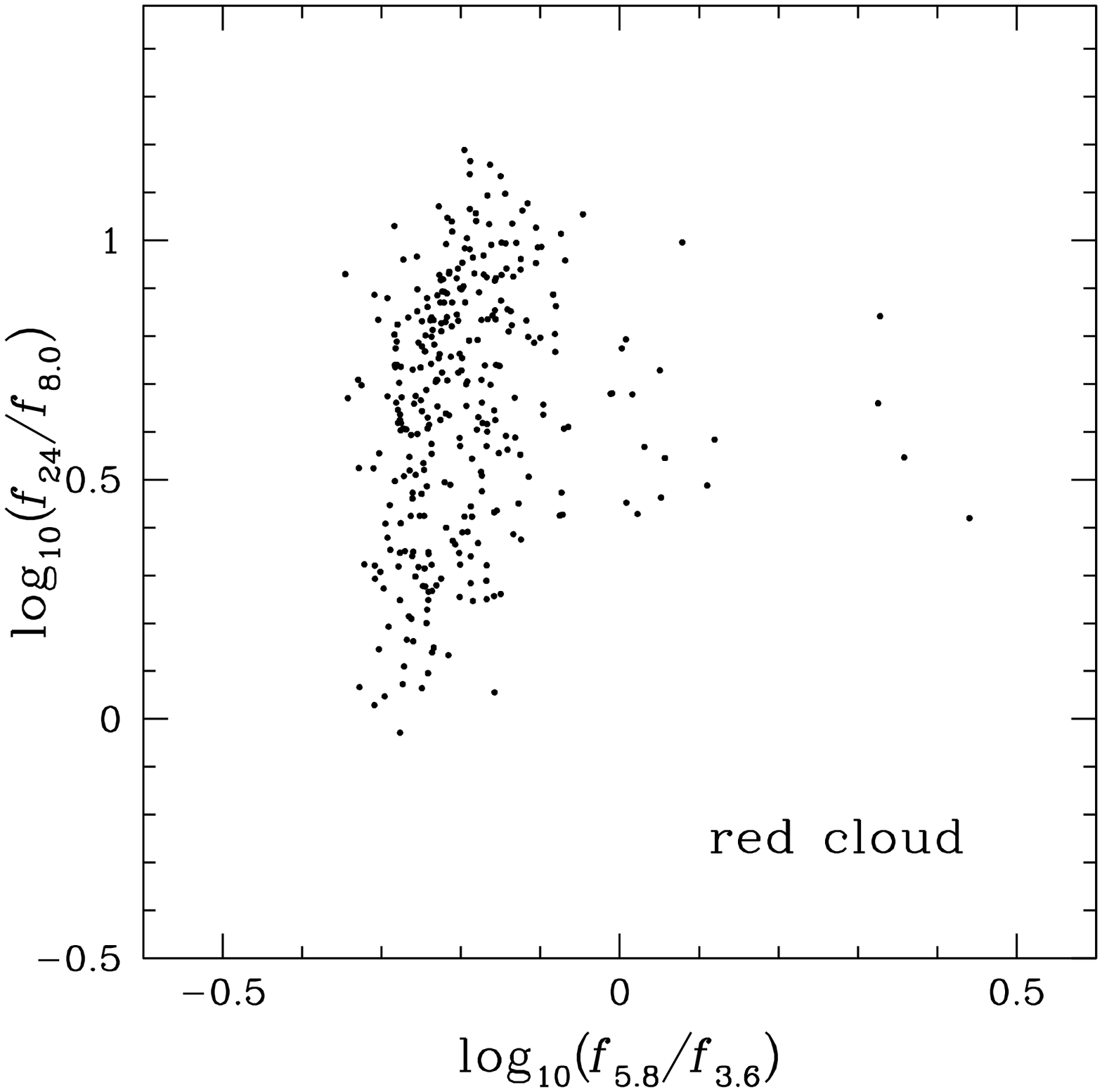} \includegraphics[height=0.7\columnwidth]{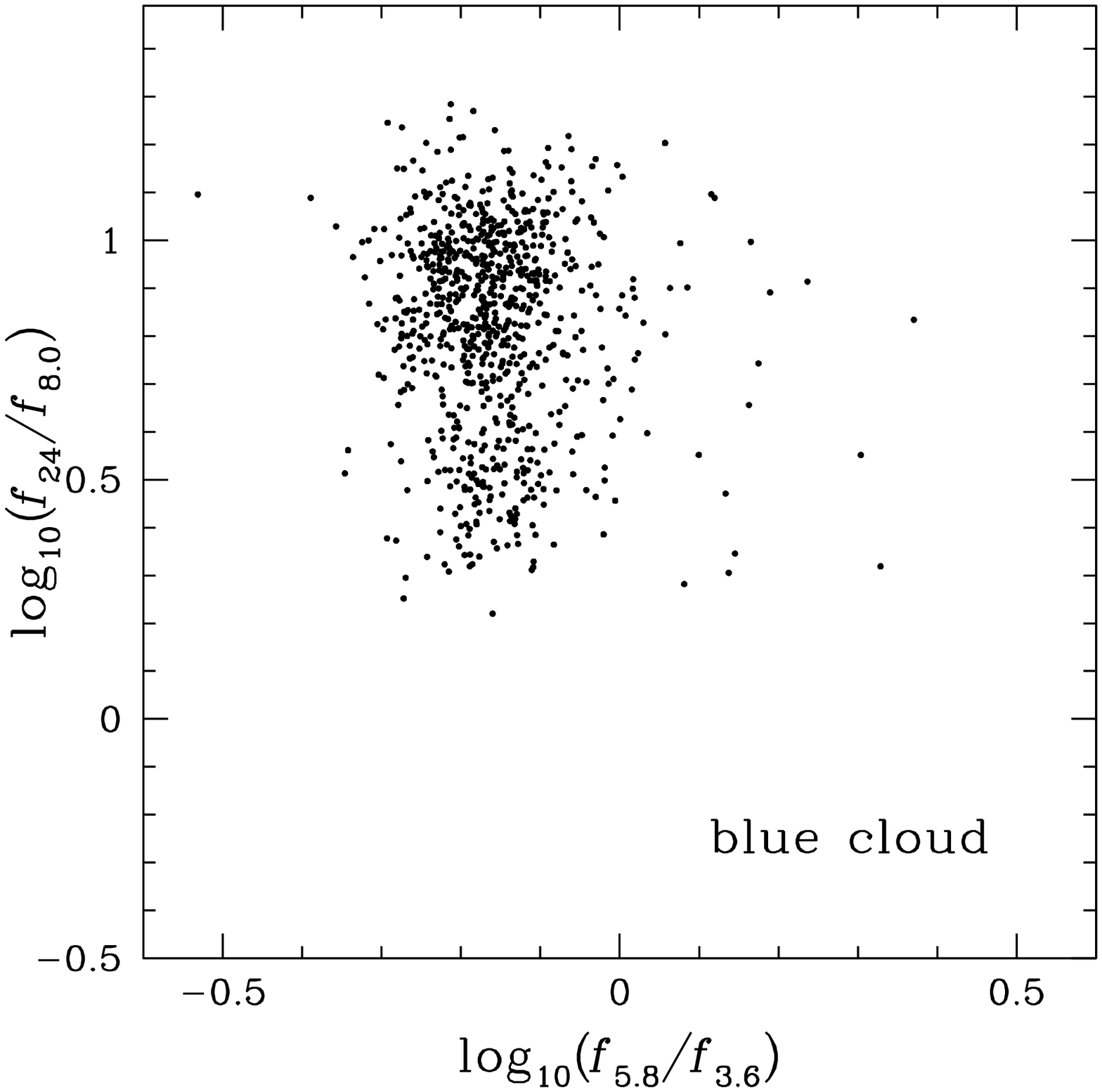} \includegraphics[height=0.7\columnwidth]{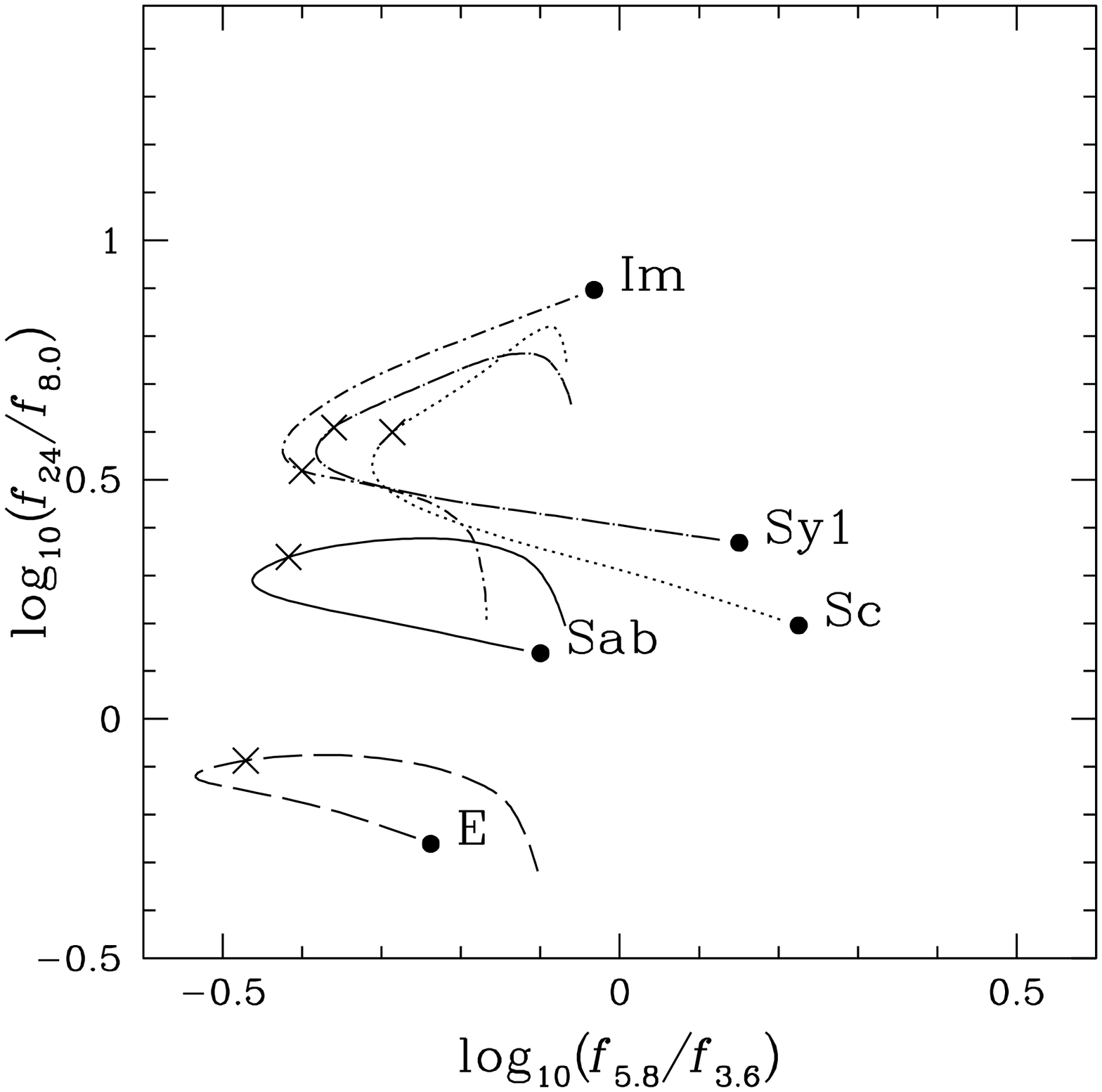}}
\end{center}
\caption{Mid-IR  colour-colour  plot.  For  clarity we  use  different
  panels.  Left: red  cloud galaxies  in the  redshift  range $0.4-0.9$.
  Middle: blue cloud  galaxies in the same redshift  interval. In both
  panels we only  plot sources detected in all  4 IRAC/MIPS bands used
  to  estimate colours.  Right:  expected tracks  of different  galaxy
  types in the range $z=0-1$ using observed mid-IR SEDs from the SINGS
  described by  Dale et al.  (2007).   NGC\,1097 Sy1:  long-dashed
  dotted; NGC\,5408 Im:  short-dashed dotted; Sc
  spiral  galaxy NGC\,628: dotted;  Sab spiral  NGC\,4450: continuous;
  elliptical galaxy NGC\,584: long  dashed. The dots mark the
  position of $z=0$ and the crosses correspond to $z=0.4$. 
  }\label{fig_lacy24}
\end{figure*}

\section{Multiwavelength observations}
 
Details about the  AEGIS datasets can be found in  Davis et al. (2007).
In  this paper  we  use (i)  the  deep {\it  Chandra} observations  to
explore the  X-ray properties  of optical galaxies  with spectroscopic
redshifts from the DEEP2 survey,  (ii) the {\it Spitzer} IRAC and MIPS
$\rm 24\,\mu m$ data to  select subsamples based on mid-IR colour, and
(iii)  the high  resolution  {\it HST}/ACS  imaging for  morphological
studies.

The X-ray data are from the {\it Chandra} survey of the Extended Groth
Strip (EGS). The observations consist of 8 ACIS-I pointings, each with
a total integration time of about  200\,ks split in at least 3 shorter
exposures obtained  at different  epochs.  The data  reduction, source
detection and flux estimation are described in detail by Nandra et al.
(2007, in preparation) and are based on methods presented by Nandra et
al.  (2005).   Briefly, standard reduction  steps are taken  using the
{\sc  ciao} version  3.2 data  analysis software.   After  merging the
individual  observations  into a  single  event  file, we  constructed
images in four energy  bands 0.5-7.0\,keV (full), 0.5-2.0\,keV (soft),
2.0-7.0\,keV (hard) and 4.0-7.0\,keV (ultra-hard).  The count rates in
the above  energy intervals  are converted to  fluxes in  the standard
bands 0.5-10,  0.5-2, 2-10 and 5-10\,keV,  respectively.  The limiting
flux in each of these bands  is estimated to be $3.5 \times 10^{-16}$,
$1.1  \times 10^{-16}$,  $8.2 \times  10^{-16}$, and  $\rm  1.4 \times
10^{-15}  \,  erg \,  s^{-1}  \,  cm^{-2}$,  respectively.  The  X-ray
catalogue comprises  a total of 1318 sources  over $\rm 0.63\,deg^{2}$
to a Poisson detection probability threshold of $4 \times 10^{-6}$. At
$z\approx1$  the  2-10\,keV  limit   above  corresponds  to  an  X-ray
luminosity   of  about   $3\times10^{42}   \rm  \,   erg  \,   s^{-1}$
($\Gamma=1.4$),   i.e.   typical   of  Seyferts.    For   the  optical
identification  we  use  the  DEEP2  photometric  catalogues  and  the
Likelihood Ratio ($LR$) method (e.g.  Brusa et al.  2007).  A total of
903  sources  have counterparts  to  $R_{AB}=25$\,mag ($LR>0.7$),  the
approximate limit of the DEEP2 photometric survey.

The DEEP2 redshift  survey uses the DEIMOS spectrograph  (Faber et al.
2003) on the 10\,m Keck-II  telescope to obtain redshifts for galaxies
to $R_{AB}  = 24.1$\,mag.  The  observational setup uses  a moderately
high  resolution grating ($R\approx5000$),  which provides  a velocity
accuracy  of  $\rm  30\,km\,s^{-1}$   and  a  wavelength  coverage  of
6500--9100\,\AA.   This spectral window  allows the  identification of
the strong [O\,II] doublet 3727\AA\, emission line to $z<1.4$.  We use
DEEP2  galaxies with  redshift  determinations secure  at the  $>90\%$
confidence level (quality flag $Q \ge 3$; Davis et al. 2007).

The {\it Spitzer}  IRAC and MIPS-$24\rm \mu m$  observations cover the
central $\rm 10  \times 120 \, arcmin^2$ subregion  of the EGS (Barmby
et al.  2006).  The integration  time is about 2.7\,hr and 1200\,s per
pointing  for  the  IRAC  and MIPS  observations,  respectively.   The
$5\sigma$  flux  density limit  for  point  sources  is 0.9,  5.8  and
83\,$\rm \mu Jy$ at 3.6, 8.0 and 24\,$\rm \mu m$, respectively.  There
are about 57\,400, 13\,600 and  6\,300 detections to the limits above.
These  sources  are matched  to  the  DEEP2  photometric catalogue  by
finding the nearest neighbour within a search radius of 1.0\,arcsec.

{\it HST}  images of the EGS  were taken with the  Advanced Camera for
Surveys (ACS) in the $V$ (F606W, 2260s) and $I$ (F814W, 2100s) filters
over a  $\rm 10.1 \times 70.5  \,arcmin ^2$ strip (Lotz  et al. 2007).
The 5\,sigma  limiting magnitudes for  a point source  are $V_{F606W}=
28.14$ (AB) and $I_{F814W} =  27.52$ (AB).  For extended sources these
limits are  about 2\,mag brighter.   A total of 15\,797  galaxies were
detected to  $I_{F814W} = 25$\,mag.   These were matched to  the DEEP2
photometric catalogue using an 1.0\,arcsec matching radius.

\section{Sample selection}\label{sec_sample}

The  main sample  used in  this paper  consists of  optically selected
galaxies with DEEP2 spectroscopy and  redshifts in the interval $0.4 <
z < 0.9$.  This redshift  range is to minimise colour dependent biases
in the selection of galaxies  introduced by the magnitude limit of the
DEEP2 spectroscopic survey, $R_{AB}<24.1$\,mag. Red cloud sources drop
below  the survey limit  at lower  redshifts than  intrinsically bluer
galaxies (e.g.  Gerke et al.  2007).  This effect becomes increasingly
severe at $z  \ga 1$ as the $R$-band straddles  the rest-frame UV.  At
$z=0.9$ the  sample is complete to $M_B=-20$\,mag.  The lower redshift
cut,  $z=0.4$, is  to avoid  biases associated  with the  small volume
sampled by the AEGIS below this limit.

There are 4814 sources in the main galaxy sample of which about 70 per
cent overlap with the {\it Spitzer} IRAC and MIPS surveys of the AEGIS
and  30 per  cent  have $I$-band  morphological  information from  the
HST/ACS observations.  The morphological sample was compiled using the
criteria of  Lotz et  al. (2007) by  selecting galaxies  brighter than
$I_{F814W}  = 25$\,mag  with average  signal-to-noise ratio  per pixel
$>$2.5 and petrosian  radius $>0.3$\,arcsec. About 65 per  cent of the
sources detected  on the HST/ACS images fulfil  these criteria. Almost
all of them (97 per cent) overlap with the IRAC/MIPS area.

The  CMD   of  the   main  galaxy  sample   is  presented   in  Figure
\ref{fig_cmd}. The blue and red  clouds are defined using the relation
of Willmer  et al.   (2006).  Galaxies with  $U-B$ colours  $\pm 0.05$
about this relation define the  valley between the two clouds. Willmer
et  al. (2006)  find that  the redshift  evolution of  the  red galaxy
population $U-B$  colour is not strong  in DEEP2.  As a  result a line
with fixed  normalisation is  adequate for defining  the blue  and red
clouds over  the redshift  interval of our  sample. The  next sections
discuss the properties of galaxies in narrow slices of the CMD.  These
are  defined  to run  parallel  to the  Willmer  et  al.  (2006)  line
separating the  two clouds and  are parametrised by their  distance in
$U-B$ colour, $\rm \Delta C$, from that line.

The post-starburst sub-sample was drawn from the main  galaxy sample
following the method  described by Yan et al.   (2006, 2007 in prep.).
The  continuum-subtracted  DEEP2  optical   spectra  were fit  with  a
combination of an old  (K-component) and a young (A-component) stellar
population SED.  Post-starburst candidates are defined as systems with
$f_A>0.25$, where  $f_A$ is the  fraction of light contributed  by the
young stellar  component around 4500\AA\, in  the linear decomposition
of  the spectrum.   The equivalent  width of  $\rm H\beta$,  after the
removal  of   the  stellar  absorption,  was  then   used  to  identify
residual/ongoing  star-formation activity within  the sample.   Yan et
al.  (2006) showed that, for high redshift galaxies, this line is a more
reliable star-formation indicator than  the [OII]\,3727 line, which is
often  associated  with  AGN  activity.   Post-starbursts  (K+As)  are
defined  as systems  with little  $\rm  H\beta$ emission,  ${\rm EW  (
H\beta )} < 5 \, f_{A} -  1$ (total of 44).  Because of the wavelength
coverage of the DEEP2  spectroscopy, post-starbursts can be identified
only in the redshift slice $0.68 - 0.88$.  In this interval, the DEEP2
spectral  window  includes  the  $\rm  H\beta$ line  as  well  as  the
rest-frame wavelength  range $3900 \la \lambda \la  4900$\AA, which is
essential  for the decomposition  of the  spectrum into  an old  and a
young stellar components.  Figure  \ref{fig_cmd} shows the position of
post-starbursts in the CMD. Most of them are in the red cloud.

The  mid-IR SED  of  galaxies  results from  a  combination of  cirrus
radiation,  stellar emission, dusty  star-formation, and  possibly hot
dust associated with an AGN.  Samples selected at $24 \,\rm \mu m$ are
generally  dominated  by  dust   enshrouded  galaxies,  both  AGN  and
starbursts (e.g.   Yan et al.  2004).  In this paper, we  will use the
$\rm 24\, \mu  m$ emission to identify such systems  and to study them
separately from quiescent  galaxies.  Figure \ref{fig_lacy24} presents
the  {\it Spitzer}  IRAC/MIPS  colours of  DEEP2/AEGIS sources.   Also
shown in  this plot are the  colour tracks for  different galaxy types
using a selection of observed SEDs from the SINGS program (Dale et al.
2007).  Undoubtedly there are large intrinsic variations in the mid-IR
properties of galaxies of a given  type resulting in a range of mid-IR
colours. Nevertheless, the general trend in Figure \ref{fig_lacy24} is
for  quiescent  systems to  occupy  the lower  part  of  the plot  and
star-formation or  AGN activity to produce redder  mid-IR colours.  In
this paper we define ``$\rm 24\,\mu m$--bright'' sources as those with
$\log(f_{24}/f_{8.0})     >     0.4$.      Bluer    mid-IR     colours
($\log(f_{24}/f_{8.0})  < 0.4$)  or no  detection at  $\rm  24\,\mu m$
defines the  ``$\rm 24\,\mu m$--faint''  sample. The latter  sample is
likely to  include ``$\rm 24\,\mu m$--bright'' sources  that are below
the detection limit  of the $\rm 24\,\mu m$  observations. As expected
most  of  the   $\rm  24\,\mu  m$  selected  sources   lie  above  the
$\log(f_{24}/f_{8.0}) > 0.4$ cut in Figure \ref{fig_lacy24}. Quiescent
galaxies are nevertheless also present in that sample.

\begin{table} 
\caption{Fraction of X-ray detections in different subsamples}\label{tab_xdet} 
\begin{center} 
\scriptsize
 \begin{tabular}{l c c c }
\hline 
Sample & $N_{TOT}$ &  $N_{X}$ &  $f_X$ \\
       &          &          &  ($\%$)  \\
  (1)  &  (2)     &    (3)   &  (4)   \\
\hline

red $0.4<z<0.7$  & 548 & 36 & $6.5\pm1.1$ \\

red $0.7<z<0.9$  & 490 & 20 & $4.1\pm1.0$ \\

red post-starburst   & 34 & 5   & $14.7\pm7.0$   \\
\\
valley $0.4<z<0.7$ & 105 & 4 & $3.8\pm1.9$ \\

valley $0.7<z<0.9$  & 126 & 8 & $6.3\pm2.3$ \\

valley post-starburst  & 9 & 1   &  $11\pm11$  \\
\\

blue $0.4<z<0.7$ & 2007 & 24 & $1.2\pm0.2$ \\

blue $0.7<z<0.9$  & 1769 & 28 & $1.5\pm0.3$ \\

blue post-starburst  & 10 & 2   & $20\pm15$ \\
\hline
\end{tabular} 

\begin{list}{}{}
\item 

The columns  are: (1): Sample  definition. Note that there  is overlap
between  the red  or  the blue  cloud  and the  valley.  The  redshift
interval $0.7<z<0.9$  is comparable in  terms of selection  effects to
post-starburst galaxies; (2):  $N_{TOT}$ is  the total number  of DEEP2
galaxies  in the  sample; (3):  $N_{X}$ corresponds  to the  number of
DEEP2 galaxies with X-ray counterparts;  (4): $f_X$ is the fraction of
X-ray detections in the sample, i.e.  $f_X=N_{X}/N_{TOT}$.  The errors
are estimated assuming Poisson statistics.

\end{list}

\end{center}
\end{table}

\begin{figure}
\begin{center}
 \rotatebox{0}{\includegraphics[height=0.9\columnwidth]{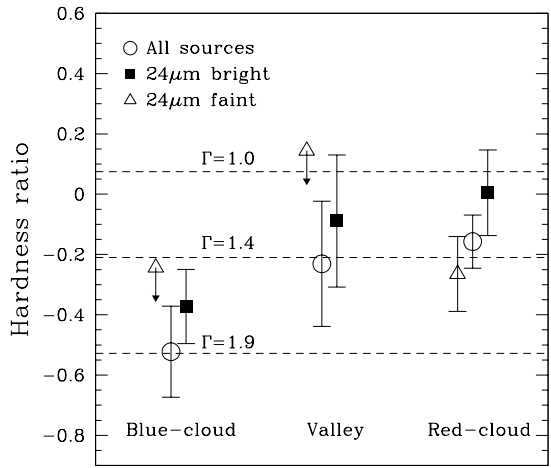}}
\end{center}
\caption{  Mean hardness  ratio for  galaxies in  the blue/red  clouds
  and the
  valley. Within  each of  these 3  groups, open
  circles correspond to all sources in the group, squares are for $\rm
  24\, \mu m$ bright galaxies, and triangles represent $\rm 24\, \mu m$
  faint  systems.   The  hardness  ratio uncertainties  are  estimated
  assuming Poisson  statistics.  In  the case of  no detection  in the
  hard  band, the  $3\sigma$  upper  limit in  the  hardness ratio  is
  plotted. The  horizontal lines  correspond to the  expected hardness
  ratio of a power-law X-ray  spectrum with spectral index from top to
  bottom  $\Gamma=1.0$,  1.4   and  1.9,  respectively.   Despite  the
  error bars, there is evidence for hardening of the mean X-ray spectrum
  from the  blue to the  red cloud.  $\rm 24\,  \mu m$ bright
  sources have harder spectra in each group.  }\label{fig_clouds}
\end{figure}

\begin{figure}
\begin{center}
 \rotatebox{0}{\includegraphics[height=0.9\columnwidth]{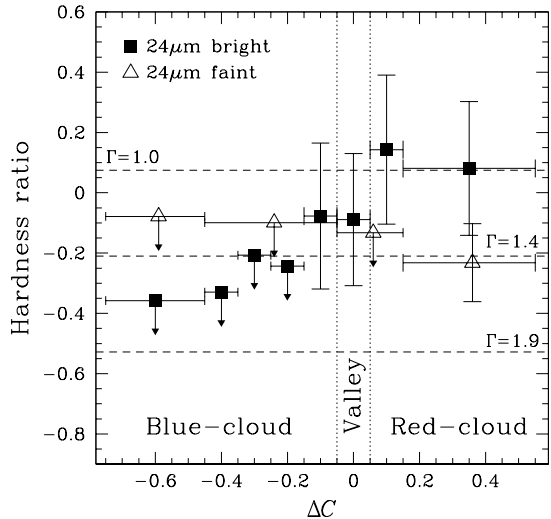}}
\end{center}
\caption{ Mean  hardness ratio estimated by  stacking optical galaxies
  within different  slices of the CMD.  $\Delta C$ is  defined as the
  difference between the colour of  the galaxy and the line separating
  the  blue   from  the  red   clouds  (Willmer  et  al.    2006;  see
  Fig. \ref{fig_cmd}).   The colour slices  are parallel to  this line
  and their position in the CMD depends on $M_B$.  The vertical dotted
  lines define  the valley  between the red  and the blue  clouds. The
  horizontal dashed lines are  the same as in Figure
  \ref{fig_clouds}. The line of  $\Gamma=1.4$ corresponds to the mean
  hardness ratio of the X-ray detected AGN.
  $\rm 24\, \mu m$-bright and $\rm 24\, \mu m$-faint systems are shown
  with  the squares  and  the triangles  respectively. The  horizontal
  error bar of  each point  corresponds to the  width of the  CMD slice
  within which galaxies are stacked.  The hardness ratio uncertainties
  are  estimated  assuming Poisson  statistics.   In  the  case of  no
  detection  in  the hard  band,  the  $3\sigma$  upper limit  in  the
  hardness    ratio   is    plotted   (arrows    pointing
  downward). 
  }\label{fig_colour}
\end{figure}

\begin{figure}
\begin{center}
 \rotatebox{0}{\includegraphics[height=0.9\columnwidth]{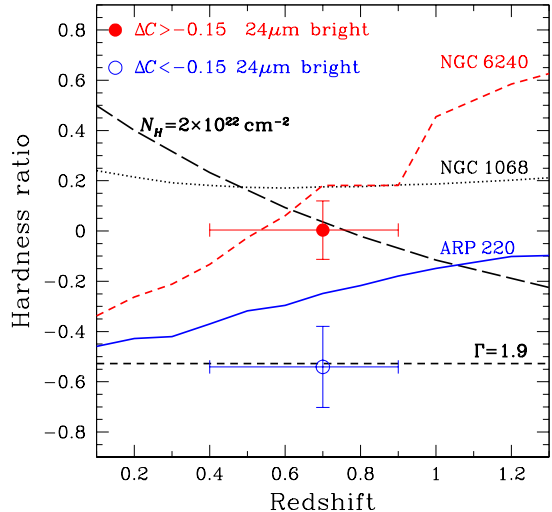}}
\end{center}
\caption{  Hardness  ratio  against  redshift.   The  curves  are  the
expected tracks  for the ARP\,220  dusty starburst (Ptak et al. 2003),
the  Compton thick AGN NGC\,1068  (Matt et al.  1999) and NGC\,6240 
(Ptak et al.  2003) and a simple absorbed power-law model with
$\Gamma=1.9$ and $N_H=\rm 2 \times 10^{22} \, cm^{-2}$.  The
short-dashed line corresponds to power-law X-ray spectrum
with $\Gamma=1.9$, typical of unabsorbed AGN. The filled
(red) circle corresponds to red $\rm  24\mu   m$-bright  galaxies with
$\Delta C>-0.15$. The  open (blue) circle is
for blue $\rm 24\mu  m$-bright galaxies with $\Delta C<-0.15$.  
}\label{fig_hr}
\end{figure}

\begin{figure}
\begin{center}
 \rotatebox{0}{\includegraphics[height=0.9\columnwidth]{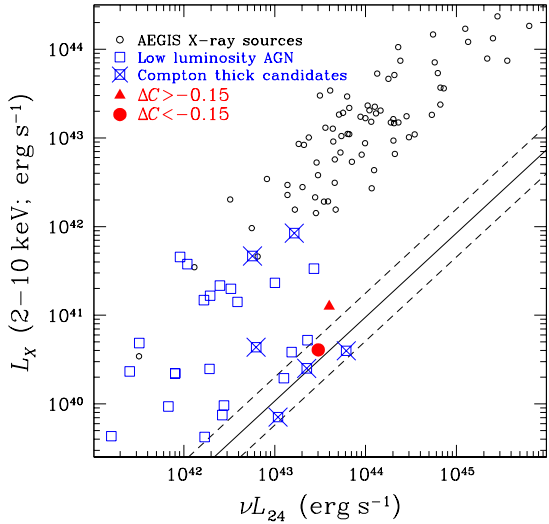}}
\end{center}
\caption{   2-10\,keV  X-ray  luminosity   against  $\rm   24\,\mu  m$
luminosity.  The  continuous line is  the $L_X - \nu  L_{24}$ relation
for star-forming  galaxies adapted from  Ranalli et al.   (2003).  The
dashed  lines correspond  to  the 1\,sigma  rms  envelope around  this
relation.  The (red) triangle corresponds to the mean $\nu L_{24}$ and
$L_X$  for  $\rm 24\,\mu  m$-bright  galaxies  with $\Delta  C>-0.15$.
Similarly,  the (red)  filled  circle is  for  $\rm 24\,\mu  m$-bright
galaxies  with optical  colours bluer  than $\Delta  C <  -0.15$.  The
uncertainties in the luminosities of these two populations are smaller
than the size of the points  and are not plotted for clarity. The open
(black) circles  are AEGIS X-ray AGN  selected in the  hard X-ray band
(2-10\,keV).   Also shown is  the local  sample of  low-luminosity AGN
(open blue squares), which  includes Compton thick candidates (crossed
blue squares),  from Terashima et al.  (2002).   The mid-IR luminosity
of  individual AEGIS X-ray  sources, $\nu  L_{24\mu m}$,  is estimated
from the $\rm 24\,\mu m$ flux density by adopting the SED of NGC\,1068
for  the k-correction.   The  conclusions are  not  sensitive to  this
assumption.  Also,  for AEGIS X-ray sources the  $L_X(\rm 2-10 \,keV)$
is calculated  from the 2-10\,keV observed-frame flux  using a typical
intrinsic  AGN spectrum  of $\Gamma  = 1.9$  (Nandra \&  Pounds 1994).
This effectively produces absorption-corrected fluxes for sources with
column densities  $ N_H <  10^{23} \rm \,  cm^{-2}$ at $z  \approx 1$.
For  the  Terashima et  al.   (2002)  low-luminosity  AGN we  use  the
obscuration corrected  luminosity listed  by these authors,  except in
the  case of Compton  thick candidates,  identified by  the equivalent
width of  the FeK line, adopting the  criterion $\rm EW >  900 \, eV$.
The $\rm 24\mu  m$ luminosity of the Terashima  et al.  (2002) sources
is estimated  using the IRAS $\rm  25\,\mu m$ flux  density.  The mean
$\rm 24\mu  m$ luminosity for the  subsamples used in  the stacking is
estimated by taking the average of the $\nu L_{24\mu m}$ of individual
sources.  The  mean $L_X ( \rm 2  - 10 \,keV)$ is  estimated using the
X-ray flux determined  by the stacking analysis and  the mean redshift
of the sources in each subsample.  }\label{fig_lxl24}
\end{figure}

\section{The X-ray stacking method}\label{sec_stacking}

The  mean  X-ray properties  of  sources  which  are not  individually
detected to the limit of  the X-ray survey are explored using stacking
analysis (e.g.   Nandra et al.   2002; Laird et  al. 2005). We  used a
fixed radius aperture  to extract and to sum the  X-ray photons at the
positions of optically selected  galaxies.  Sources that are separated
from an X-ray  detection by less than 1.5 times the  local 90 per cent
Encircled  Energy   Fraction  (EEF)  radius  are   excluded  from  the
stacking. This  is to avoid  contamination of the stacked  signal from
photons associated with the Point Spread Function (PSF) wings of X-ray
detections. Also, in order to  exclude regions where the {\it Chandra}
PSF is  large and the  sensitivity is low,  we do not  extract photons
from  pointings  where  a  particular  source  is  located  more  than
9\,arcmin away from  the centre of the detector.   This off-axis angle
cutoff is chosen to maximise  the stacked signal, although our results
are not particularly sensitive  to this parameter.  For the extraction
radius, we  experimented with apertures in the  range 1--3\,arcsec and
selected  2\,arcsec   as  the   optimal  radius  that   maximises  the
signal-to-noise ratio of the stacked X-ray photons.

To assess  the significance  of the stacked  signal, we  estimated the
local background of a particular source by averaging the X-ray photons
within a 50\,arcsec  radius (100 pixels) and then  scaling to the area
of the  extraction aperture. When determining the  local background we
clipped  regions around  X-ray  detections using  a  radius 1.5  times
larger than  the 90 per cent  EEF. The significance (in  sigma) of the
stacked signal is estimated by $(T-B)/\sqrt{B}$, where $T$ and $B$ are
the total  (source +  background) and background  counts respectively.
The 2\,arcsec radius includes only a fraction of the source photons at
each  position. We  account  for the  remaining  flux when  estimating
count-rates  and  fluxes  by   applying  a  mean  aperture  correction
determined by averaging the exposure-time weighted PSF corrections for
individual sources.

\section{Results}\label{results}

\subsection{X-ray properties of galaxies in the red cloud and in the transition zone} 

Table \ref{tab_xdet} summarises the fraction of X-ray detections among
galaxy samples  selected at several  positions in the CMD.   The X-ray
identification  rate is  higher in  the red  cloud, in  agreement with
recent studies  on the  host galaxy properties  of X-ray  selected AGN
(Nandra et al.  2007).

In  this   section,  we  search   for  evidence  of   obscured  and/or
low-luminosity systems below the X-ray detection threshold by stacking
optically  selected galaxies  in  different regions  of  the CMD.   Of
particular interest is the valley,  where AGN are suggested to play an
important  role  in  galaxy  evolution.   The  results  for  different
subsamples are summarised in  Table \ref{tab_stack} and are plotted in
Figure  \ref{fig_clouds}. There  is  some evidence  for a  progressive
hardening of the stacked signal from  the blue to the red cloud.  This
is  further demonstrated  in Figure  \ref{fig_colour} which  plots the
hardness ratio  of galaxies  in CMD slices  which run parallel  to the
Willmer et al.   (2006) line separating the blue  from the red clouds.
Although  the errorbars  of individual  points are  large, there  is a
systematic  trend whereby  the mean  X-ray spectrum  of $\rm  24\, \mu
m$-bright sources  becomes harder, reaching $\Gamma  \approx 1.2$, for
$\Delta C>-0.15$, i.e.   for sources around the valley  and in the red
cloud.  Bluer  galaxies have  hardness ratio upper  limits ($3\sigma$)
consistent  with   softer  mean  X-ray   spectra,  $\Gamma>1.4$.   The
corresponding upper limits for the $\rm 24\, \mu m$-faint subsample do
not  provide  strong constraints,  but  also  suggest relatively  soft
spectral properties.  In any case,  there is no evidence that the mean
hardness ratio  of this  population increases with  $\Delta C$  at the
same level as for $\rm 24\, \mu m$-bright galaxies.

In  order  to  improve the  statistics  we  split  the $\rm  24\,  \mu
m$-bright sample  at $\Delta  C=-0.15$ and stacked  separately sources
bluer/redder than  this limit.  The  resulting mean hardness  ratio is
plotted as  a function of  redshift in Figure \ref{fig_hr}.   The mean
hardness ratio of the $\rm 24\,\mu m$-bright population with $\Delta C
>-0.15$ is harder  than the ARP\,220 starburst template  and in better
agreement  with the obscured  AGN models.   In contrast,  the hardness
ratio  of galaxies  bluer than  $\Delta C  <-0.15$ is  consistent with
star-formation (i.e.  X-ray binaries  and hot gas), although we cannot
exclude the possibility of low-luminosity unobscured AGN.

The obscured AGN interpretation  for the $24\,\mu m$-bright population
with $\Delta  C >-0.15$ is  also supported by  Figure \ref{fig_lxl24},
which plots the  2-10\,keV X-ray luminosity against the  $\rm 24 \,\mu
m$  luminosity, in  comparison  with the  relation  between these  two
quantities  for  star-forming galaxies  adapted  from  Ranalli et  al.
(2003).   Sources with  $\Delta C  >-0.15$  in this  figure are  X-ray
luminous  compared to  this  relation, suggesting  that their  stacked
X-ray  signal is  dominated by  AGN.  The  same conclusion  applies if
galaxies  are split  into finer  colour bins,  e.g.  upper  blue cloud
($-0.15 < \Delta C < -0.05$),  valley ($-0.05 < \Delta C < -0.05$) and
red cloud ($\Delta C >  +0.05$).  These subsamples have mean X-ray and
$24\,\mu  m$ luminosities  similar to  the overall  $\Delta  C >-0.15$
population  and therefore  are  also X-ray  luminous  compared to  the
expectation from star-formation.  These results are in contrast to the
mean X-ray luminosity of $24\, \mu m$-bright galaxies with $\Delta C <
-0.15$, which  is consistent  with the star-formation  relation. These
suggests that X-ray  binaries and hot gas dominate  the X-ray emission
of this population.

\subsection{X-ray properties of post-starbursts}

An interesting trend in Table \ref{tab_xdet} is the higher fraction of
X-ray  sources in  post-starbursts. In  the red  cloud  in particular,
where the majority of post-starbursts  are found, 15 per cent of these
systems  are associated  with X-ray  sources.  In  contrast  the X-ray
identification rate is only 2  per cent for the overall optical galaxy
population and  about 4 per  cent for galaxies  in the red  cloud.  In
order  to assess  the significance  of  the excess,  we resampled  the
optical galaxy  population to construct subsamples with  size equal to
the  number of red-cloud  post-starbursts and  with similar  $M_B$ and
$U-B$  distributions. The  fraction of  X-ray identifications  in each
subsample is registered and the experiment was repeated 10\,000 times.
The X-ray detection  rate of the random subsamples  is lower than that
of post-starbursts in  98 per cent of the  experiments.  The excess of
X-ray sources  in this population  is therefore significant at  the 98
per cent  level.  Post-starbursts are also  a non-negligible component
of  the  X-ray population.   In  the  redshift  interval $0.68  \la  z
\la0.88$, $21\pm10$\%  (5/24) of  the X-ray sources  in the  red cloud
have post-starburst  optical spectra.  For comparison  the fraction of
red cloud  galaxies that are  post-starbursts is 7 per  cent (34/490).
These   results,  although   limited  by   small   number  statistics,
tentatively suggest a link between AGN and the post-starburst stage of
galaxy evolution at $z \approx 0.8$. The link is also supported by the
mean  stacked  X-ray properties  of  post-starbursts not  individually
detected   to  the  limit   of  AEGIS   {\it  Chandra}   survey.   The
post-starburst population is marginally  detected in both the soft and
the hard spectral  bands at a significance level  $\ga 3\sigma$ (Table
\ref{tab_stack},  column  7).   Although  the stacked  signal  is  not
dominated by a single source just below the X-ray detection threshold,
small  number  statistics  are   a  concern.   The  results  in  Table
\ref{tab_stack},  taken at  face  value, are  consistent with  $\Gamma
\approx  0.7$.   This  is  much  harder than  the  X-ray  spectrum  of
star-forming galaxies (Figure \ref{fig_hr}) or unobscured AGN ($\Gamma
\approx  1.9$; Nandra \&  Pounds 1994)  suggesting that  absorption is
suppressing the soft  X-ray emission from the central  engine in these
systems.

\begin{figure*}
\begin{center}
\rotatebox{0}{\includegraphics[height=1\columnwidth]{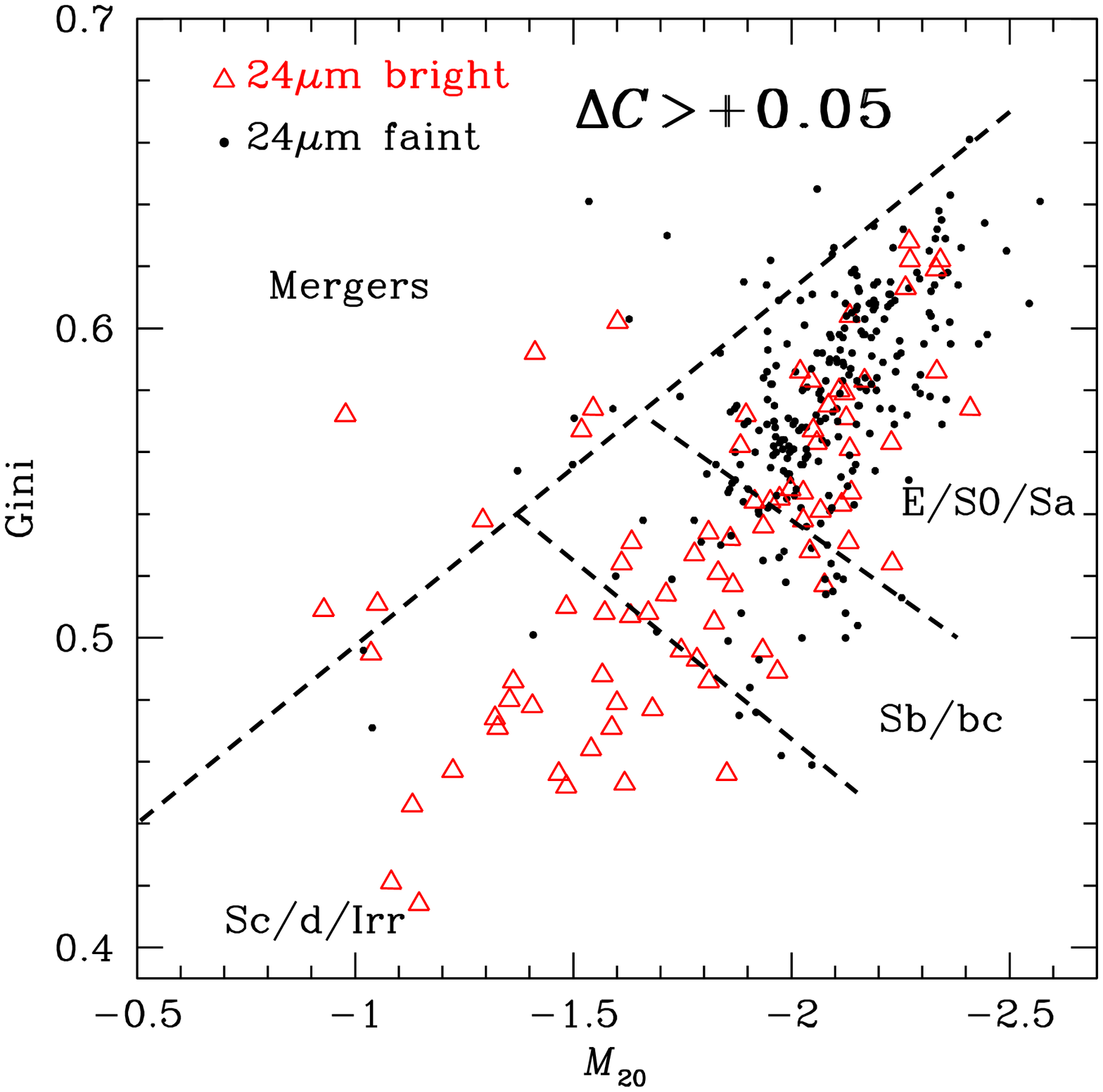}
 \includegraphics[height=1\columnwidth]{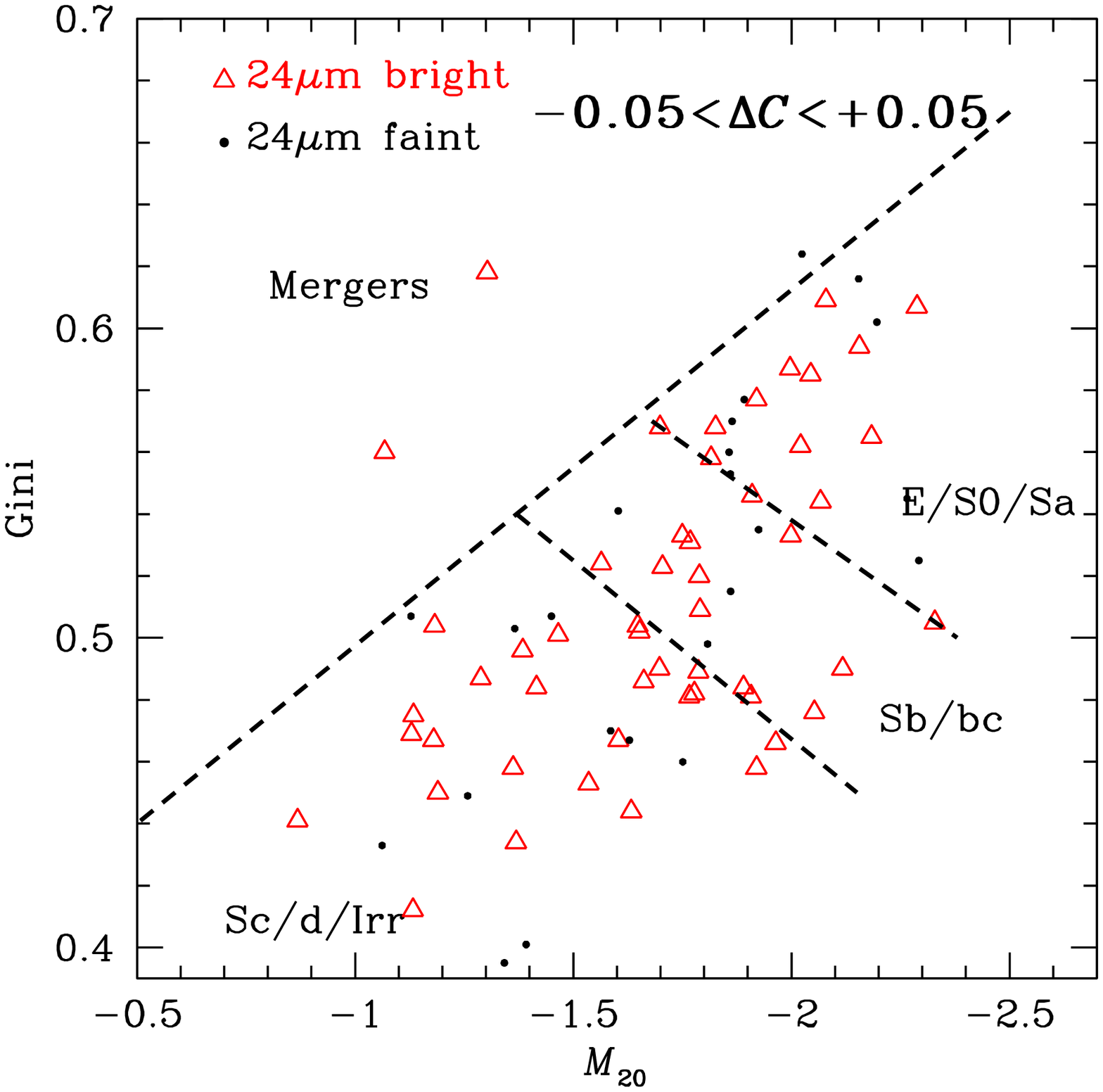}             }
\rotatebox{0}{\includegraphics[height=1\columnwidth]{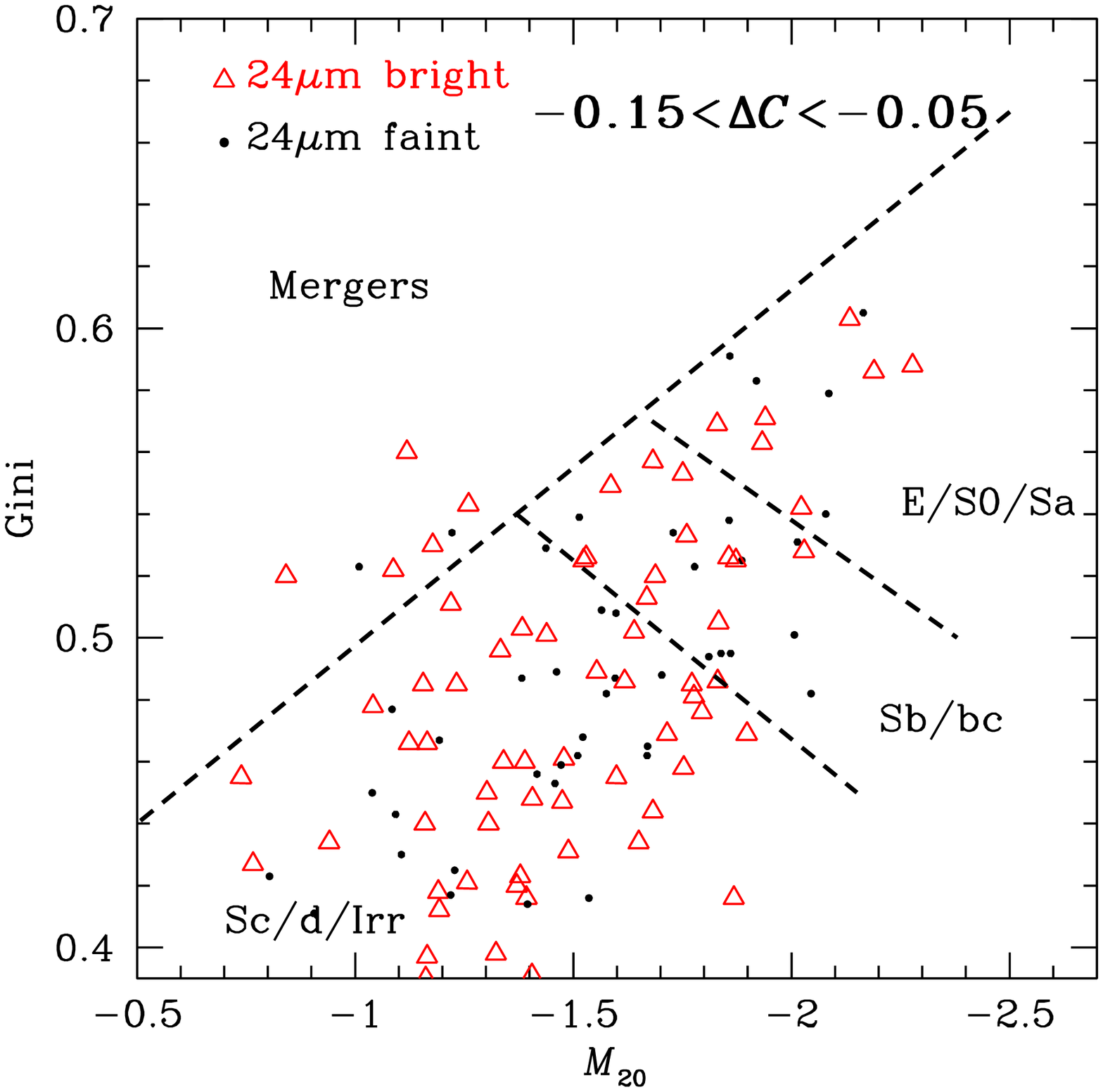}
 \includegraphics[height=1\columnwidth]{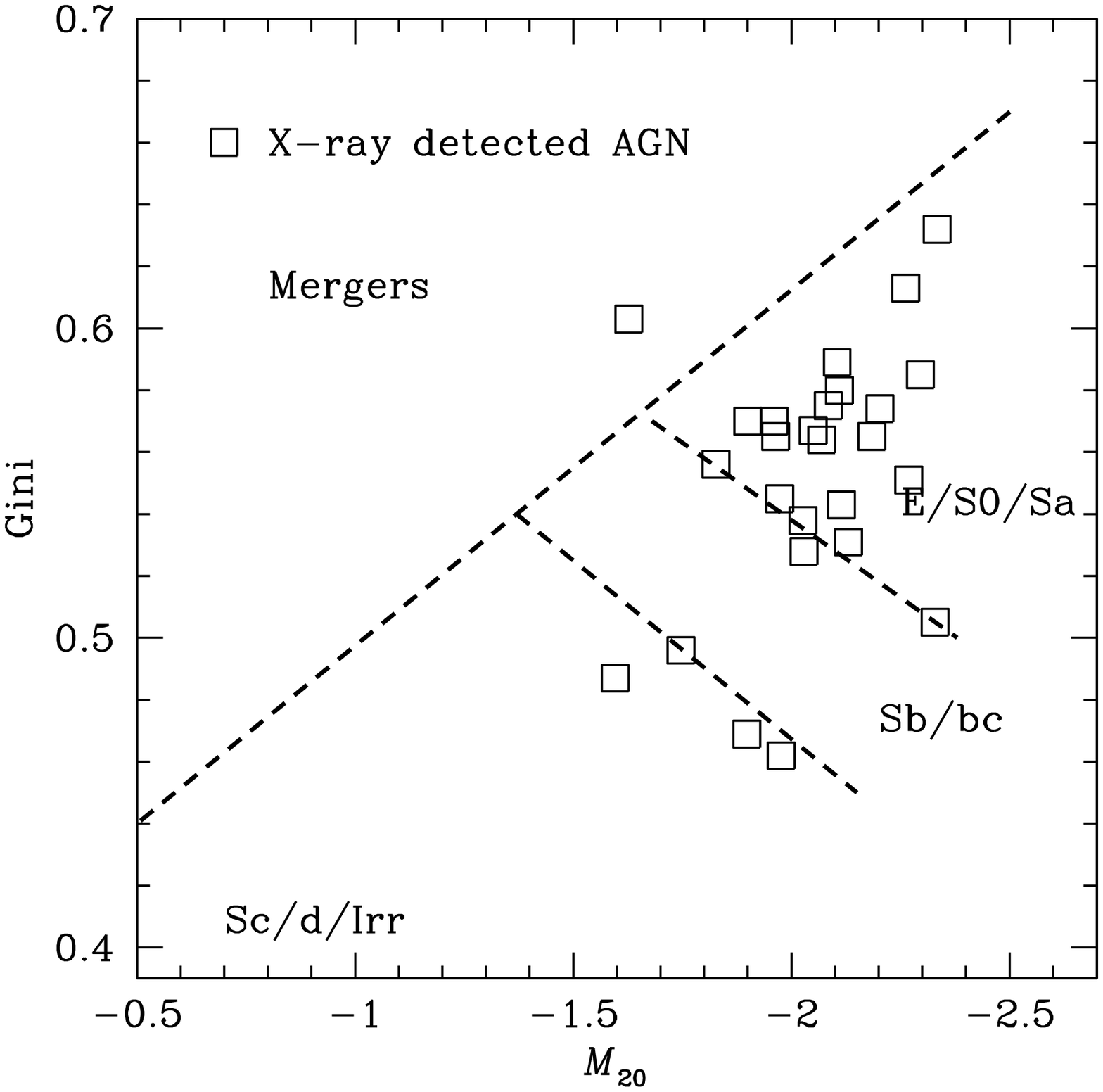} }

\end{center}
\caption{  Gini  against  $M_{20}$   diagrams.   The  regions  of  the
parameter space occupied by different galaxy types are demarcated with
the dashed  lines. For clarity sources  at different parts  of the CMD
are plotted in  different panels: upper left sources  with $\Delta C >
+0.05$,  upper right  $-0.05< \Delta  C <+0.05$  (valley),  lower left
$-0.15< \Delta C <-0.05$ (upper blue cloud).  In all these panels open
triangles are $\rm  24\mu m$ bright sources and the  dots are for $\rm
24\mu  m$  faint galaxies.   In  the lower  right  panel  we show  the
position  of  X-ray  detected   AGN  on  the  Gini--$M_{20}$  diagram.
}\label{fig_morph}
\end{figure*}

\begin{table*} 
\caption{Stacking results}\label{tab_stack} 
\begin{center} 
\begin{tabular}{l c c   c c c c c  c c c}
\hline 
Sample & N & $<z>$ & band & $T$ &  $B$ &   S/N      & photon rate & HR
& $f_X$ & $\log L_X$\\
       &      &   &       &               &                &
($\sigma$) & ($10^{-9}$) &  & ($10^{-17}$)  & ($\rm erg/s$)\\        
  (1)  &  (2) & (3) & (4)  &  (5)  &    (6)   &  (7)   &  (8)  & (9) &
(10)\\

\hline
\multicolumn{11}{c}{\Large red cloud ($\Delta C>0.0$)}\\

\multirow{2}{*}{all} & \multirow{2}{*}{881} & \multirow{2}{*}{0.67} 
        & soft & 741  & 389.9 & 17.9 & $9.6\pm0.9$ & \multirow{2}{*}{$-0.16\pm0.09$} &$1.7\pm0.2$ & 40.4\\
 &   &  & hard &1178  & 892.4 &  9.6 & $7.0\pm1.1$&    & $6.4\pm1.0$   & 40.9\\ 

\multirow{2}{*}{$\rm 24\mu m$-bright} &  \multirow{2}{*}{177} & \multirow{2}{*}{0.69}  
        & soft & 169 & 80.8 & 9.8 & $10.7\pm1.9$ &  \multirow{2}{*}{$0.01\pm0.14$} & $1.9\pm0.3$ & 40.4 \\ 
&   &  & hard & 286 &187.5 & 7.2 & $10.8\pm2.4$ & & $10.7\pm2.4$  & 41.2\\

\multirow{2}{*}{$\rm 24\mu m$-faint} & \multirow{2}{*}{513} & \multirow{2}{*}{0.66}  
        & soft & 451 & 235.2 & 14.1 & $9.5\pm1.2$    & \multirow{2}{*}{$-0.27\pm0.12$} & $1.6\pm0.2$ & 40.4\\
 &   &  & hard & 676 & 537.3 &  6.0 & $5.5\pm1.4$    & & $4.7\pm1.2$  & 40.8\\ 

\\

\multicolumn{11}{c}{\Large valley ($-0.05<\Delta C<+0.05$)}\\

\multirow{2}{*}{all} & \multirow{2}{*}{188} & \multirow{2}{*}{0.68} 
        & soft & 155  &  81.8 & 8.1 & $9.5\pm2.0$ & \multirow{2}{*}{$-0.23\pm0.21$} & $1.6\pm0.3$ & 40.4\\
     &   &  & hard & 238  & 187.1 & 3.7 & $6.0\pm2.4$ & & $5.3\pm2.1$ &  40.9\\ 

\multirow{2}{*}{$\rm 24\mu m$-bright} &  \multirow{2}{*}{89} & \multirow{2}{*}{0.70}  
        & soft &  85 & 41.0 & 6.9 & $11.4\pm2.9$ & \multirow{2}{*}{$-0.09\pm0.22$} & $2.0\pm0.5$ & 40.5\\
    &   &  & hard & 113 & 92.3 & 4.2 & $9.5\pm3.5$  & & $9.0\pm3.3$ & 41.1\\ 

\multirow{2}{*}{$\rm 24\mu m$-faint} & \multirow{2}{*}{61} & \multirow{2}{*}{0.68}  
        & soft & 42 & 26.1 & 3.1 & $5.8\pm3.0$    & \multirow{2}{*}{$<$0.15} & $1.0\pm0.5$ & 40.3\\
  &   &  & hard & 61 & 60.7 & 0.0  & $<$7.8    & & $<$6.9 & $<$41.1\\ 

\\

\multicolumn{11}{c}{\Large blue cloud ($\Delta C<0.0$)}\\

\multirow{2}{*}{all} & \multirow{2}{*}{3331} & \multirow{2}{*}{0.66} 
        & soft & 2009  & 1469.1 & 14.1 & $3.9\pm0.4$ & \multirow{2}{*}{$-0.52\pm0.15$} & $0.6\pm0.1$ & 40.0\\
 &   &  & hard & 3565  & 3375.7 &  3.3  & $1.2\pm0.5$ & & $0.9\pm0.4$ & 40.2\\ 

\multirow{2}{*}{$\rm 24\mu m$-bright} &  \multirow{2}{*}{572} & \multirow{2}{*}{0.68}  
        & soft & 538 & 269.0  & 15.8 & $9.7\pm1.1$ & \multirow{2}{*}{$-0.37\pm0.12$} & $1.6\pm0.2$ & 40.4\\ 
 &   &  & hard & 745 & 613.6  &  5.3 & $4.4\pm1.2$ & & $3.6\pm1.0$  & 40.8\\

\multirow{2}{*}{$\rm 24\mu m$-faint} & \multirow{2}{*}{1811} & \multirow{2}{*}{0.66}  
        & soft & 1018 & 823.7 & 6.8 & $2.4\pm0.5$    & \multirow{2}{*}{$<-0.24$} & $0.4\pm0.1$ & 39.8\\
 &   &  & hard & 1910 & 1897.2 & 0.3  & $<$1.4   & & $<$1.2 & $<$40.3\\ 

\\

\multicolumn{11}{c}{\Large $\Delta C>-0.15$}\\

\multirow{2}{*}{all} & \multirow{2}{*}{1270} & \multirow{2}{*}{0.67} 
        & soft & 1017  & 560.5 & 19.3 & $8.7\pm0.7$ & \multirow{2}{*}{$-0.17\pm0.08$} & $1.5\pm0.1$ & 40.3\\
 &   &  & hard & 1643  & 1279.8 &  10.2  & $6.2\pm0.9$ & & $5.7\pm0.8$ & 40.9\\ 

\multirow{2}{*}{$\rm 24\mu m$-bright} &  \multirow{2}{*}{332} & \multirow{2}{*}{0.69}  
        & soft & 300 & 153.6  & 11.8 & $9.5\pm1.4$ & \multirow{2}{*}{$+0.00\pm0.12$} & $1.7\pm0.2$ & 40.4\\ 
 &   &  & hard & 516 & 352.7  &  8.7 & $9.5\pm1.7$ & & $9.4\pm1.7$  & 41.1\\

\multirow{2}{*}{$\rm 24\mu m$-faint} & \multirow{2}{*}{648} & \multirow{2}{*}{0.66}  
        & soft & 531 & 294.4 & 13.8 & $8.2\pm1.0$    & \multirow{2}{*}{$-0.31\pm0.13$} & $1.4\pm0.2$ & 40.3\\
 &   &  & hard & 810 & 671 & 5.3  & $4.35\pm1.2$   & & $3.7\pm1.0$ & 40.8\\

\\

\multicolumn{11}{c}{\Large $\Delta C<-0.15$}\\

\multirow{2}{*}{all} & \multirow{2}{*}{2942} & \multirow{2}{*}{0.66} 
        & soft & 1733  & 1298.5 & 12.1  & $3.5\pm0.5$ & \multirow{2}{*}{$<-0.49$} & $0.6\pm0.1$ & 40.0\\
 &   &  & hard & 3100  & 2988.3 &  2.0  & $<1.2$ & & $<0.9$ & $<$40.2\\ 

\multirow{2}{*}{$\rm 24\mu m$-bright} &  \multirow{2}{*}{417} & \multirow{2}{*}{0.67}  
        & soft & 397 & 196.2  & 14.3 & $10.3\pm1.3$ & \multirow{2}{*}{$-0.54\pm0.16$} & $1.7\pm0.2$ & 40.5\\ 
 &   &  & hard & 515 & 448.4  &  3.2 & $3.1\pm1.4$ & & $2.4\pm1.1$  & 40.6\\

\multirow{2}{*}{$\rm 24\mu m$-faint} & \multirow{2}{*}{1676} & \multirow{2}{*}{0.66}  
        & soft & 938 & 764.6 & 6.3 & $2.3\pm0.5$    & \multirow{2}{*}{$<-0.21$} & $0.4\pm0.1$ & 39.8\\
 &   &  & hard & 1776 & 1763 & 0.3  & $<1.5$   & & $<$1.2 & $<$40.3\\

\\
\multirow{2}{*}{post-starburst} & \multirow{2}{*}{26} & \multirow{2}{*}{0.76}  
        & soft & 20 & 13.7 & 1.7 & $<$7.5    & \multirow{2}{*}{$>+0.35$} & $<$1.4 &$<$40.3 \\
 &   &  & hard & 44 & 25.9 & 3.6  & $15.5\pm7.1$    & & $17.8\pm8.2$ & 41.4\\ 

\\

\hline

\end{tabular} 

\begin{list}{}{}
\item

The  columns are:  (1): Sample  definition. The  redshift  interval is
$0.4-0.9$ for  all samples except for post-starbursts;  (2): number of
sources used for  stacking after excluding galaxies that  lie close to
or associated with X-ray detections; (3): mean redshift of the sample;
(4): spectral band where  X-ray photons are stacked: soft: 0.5--2\,keV
and  hard: 2--7\,keV; (5):  Total counts  (source$+$background) within
the extraction  radius; (6):  Background counts; (7):  significance of
the detected  signal estimated from the  relation $(T-B)/\sqrt{B}$ and
expressed in background standard deviations; (8): Photon count rate in
units of $\rm 10^{-9}\,s^{-1}$  corrected for aperture effects and the
{\it  Chandra} response. In  the case  of non-detection  the $3\sigma$
upper  limit is  listed; (9):  Hardness ratio  defined  as (H-S)/(H+S)
where H, S are the hard and soft band photon count rates respectively.
The errors are estimated assuming Poisson statistics; (10): Rest-frame
X-ray  flux in  units of  $\rm 10^{-17}\,  erg \,  s^{-1}  \, cm^{-2}$
estimated in the 0.5-2 and  2-10\,keV spectral bands adopting the mean
power-law  spectral index  consistent with  the  HR.  In  the case  of
non-detection the $3\sigma$ upper  limit is listed; (11): Logarithm of
rest-frame luminosity in units of $\rm erg \, s^{-1}$.

\end{list}

\end{center}
\end{table*}

\subsection{Optical morphology: interactions and star-formation}

In the  previous sections,  we found evidence  for AGN  activity among
$\rm  24\,\mu  m$-bright  galaxies  with  $\Delta C  >  -0.15$  (which
includes post-starbursts).  This cut  includes the reddest part of the
blue cloud, the valley, and  the red cloud.  The optical morphology of
these sources  should be a  powerful diagnostic of their  nature.  Are
they   interacting/merging  systems?   Do   they  show   evidence  for
star-formation?  For classifying galaxies into different morphological
types, we use the {\it HST}/ACS data to estimate the Gini coefficient,
which measures  the clumpiness of a  source, and the  second moment of
the brightest  20\% pixels of  the galaxy, $M_{20}$ (Lotz,  Primack \&
Madau  2004), which measures  the central  concentration of  a galaxy.
Different Hubble types are separated in the Gini--$M_{20}$ diagram and
the  morphological classification  based on  these  two non-parametric
estimators remains robust at  high redshift.  An additional parameter,
the  asymmetry, is  also used  to search  for disturbances  that might
indicate  recent/ongoing  interactions (e.g.   Abraham  et al.   1996;
Conselice et al.   2000). This parameter measures the  degree to which
the light of the galaxy is rotationally symmetric.

Figure  \ref{fig_morph}   plots  the  Gini--$M_{20}$   parameters  for
galaxies  with  $\Delta C  >  -0.15$.   While  the $24\,\mu  m$-bright
population  with $\Delta  C >  -0.15$ includes  systems  classified as
ongoing mergers,  these are about 7  per cent of  the population.  The
majority of the $24\,\mu  m$-bright sources in Figure \ref{fig_morph},
about 70 per  cent, have spiral or irregular  morphology, Sb or later.
Blue star-forming regions in these galaxies are indeed resolved by the
{\it HST}.  The  $24\,\mu m$ emission of these  systems is also likely
associated with young stars.   Moreover, visual inspection of the {\it
HST}/ACS images further shows  that many $24\,\mu m$-bright sources in
the E/S0/Sa  region of  the Gini-$M_{20}$ diagram  also have  disks in
addition to the dominant  bulge, i.e.  early type spirals.  Similarly,
visual inspection  of the  HST/ACS images of  the X-ray  detected AGN,
also plotted  in Figure \ref{fig_morph},  suggests that about  half of
these systems are associated with spirals.  The morphological evidence
above  suggests that  disks represent  a substantial  fraction  of the
$24\,\mu m$-bright galaxies with $\Delta  C > -0.15$ and about half of
the  X-ray  detected  AGN.   This  suggests that  (i)  some  level  of
star-formation  is likely  taking place  in the  host galaxy  and (ii)
recent major mergers are unlikely to have played a central role in the
evolution of these sources.

\begin{figure}
\begin{center}
 \rotatebox{0}{\includegraphics[height=0.9\columnwidth]{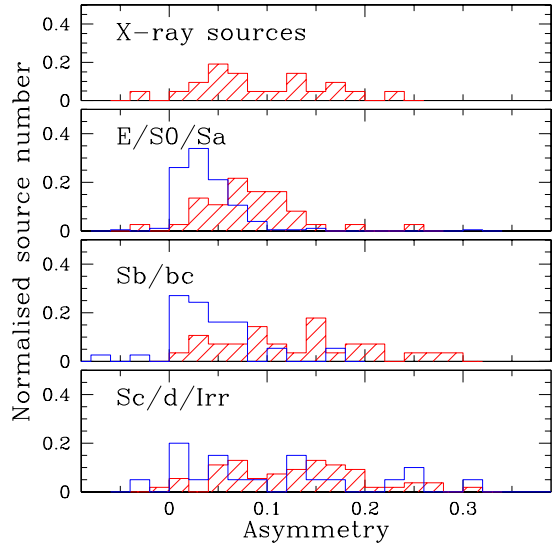}}
\end{center}
\caption{Normalised   distribution  of  the   asymmetry
  parameter for $\rm 24\,\mu m$-bright (hatched red histogram) and $\rm
  24\mu m$-faint galaxies (open blue histogram) with $\Delta C
  >-0.15$. The different 
  panels correspond to different morphological types defined using the 
  Gini-$M_{20}$ diagram shown in Figure \ref{fig_morph}.  X-ray
  detected AGN in the AEGIS are plotted in the top panel.}\label{fig_assym}
\end{figure}

Although major  mergers do  not appear to  be frequent  among $24\,\mu
m$-bright galaxies with $\Delta C > -0.15$ tidal disruptions, or minor
mergers may  play a role  in their evolution.   Figure \ref{fig_assym}
compares  the distribution  of the  asymmetry parameter  for  the $\rm
24\,\mu  m$-bright  and the  $\rm  24\,\mu  m$-faint  galaxies in  the
$\Delta  C >  -0.15$  part of  the  CMD.  The  panels  in this  figure
correspond  to  different  Hubble  types based  on  the  Gini-$M_{20}$
classification  of Figure  \ref{fig_morph}. In  the E/S0/Sa  and Sb/bc
panels,  the  $\rm  24\,\mu  m$-bright  sample  is  offset  to  higher
asymmetries compared to the  $\rm 24\,\mu m$-faint galaxies. For later
Hubble  types, Sc/d/Irr, there  is little  difference between  the two
subsamples.  A  Kolmogorov-Smirnov tests shows that  the likelihood of
the observed  differences if the  $24\,\mu m$-bright and  $\rm 24\,\mu
m$-faint  samples  were  drawn  from  the same  parent  population  is
$<10^{-6}$,  $7\times10^{-6}$, and  0.15  for the  E/S0/Sa, Sb/bc  and
Sc/d/Irr classes,  respectively.  For comparison,  X-ray detected AGN,
most of  which are in the  E/S0/Sa part of  the Gini-$M_{20}$ diagram,
are  also offset  to higher  asymmetries compared  to  early-type $\rm
24\,\mu  m$-faint   galaxies.   This  evidence   suggests  that  minor
gravitational encounters play a role in the evolution of both the $\rm
24\,\mu m$-bright systems  with early-type optical morphology (E/S0/Sa
and   Sb/bc)  and  some   of  X-ray   detected  AGN   associated  with
disks. Alternatively clumpy star-formation in the disk of galaxies can
also  produce  the  same  effect,  i.e.   higher  asymmetry  parameter
distribution.  Differentiating between  the two interpretations is not
easy, and  it is  likely that clumpy  star-formation and  minor galaxy
interactions are linked (e.g. Bell et al. 2005).

\begin{figure}
\begin{center}
 \rotatebox{0}{\includegraphics[height=0.9\columnwidth]{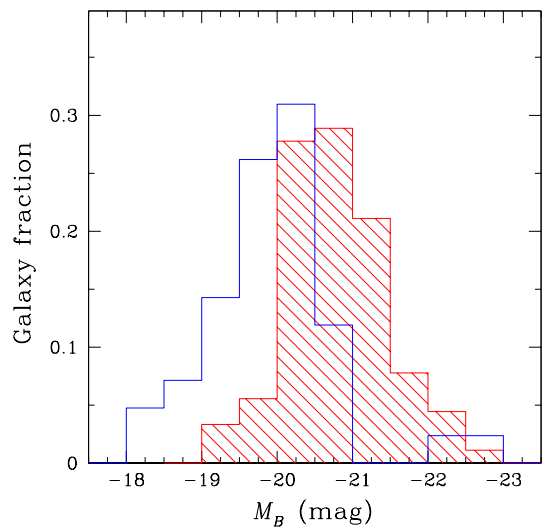}}
\end{center}
\caption{Normalised   distribution  of  the   absolute $B$-band
  magnitude for $\rm 24\,\mu m$-bright (hatched red histogram) and $\rm
  24\mu m$ faint galaxies (open blue histogram) with late-type
  optical morphology (Sc/d/Irr) and $\Delta C>-0.15$. 
}\label{fig_hist_mb}
\end{figure}

For  late   type  galaxies   (Sc/d/Irr)  there  is   no  statistically
significant difference in the asymmetry parameter distribution between
$\rm  24\,\mu m$-bright/faint sources.   The two  populations however,
have  distinct $M_B$  distributions, suggesting  differences  in their
stellar  mass.   This  is  shown in  Figure  \ref{fig_hist_mb},  where
late-type  $24\,\mu m$-bright galaxies  with $\Delta  C >  -0.15$ have
mean absolute  $B$-band magnitude  $M_B \approx -20.8$\,mag.   This is
more luminous  than the average $M_B \approx  -20.0$\,mag for $24\,\mu
m$-faint sources  of similar morphological type.   The two populations
also have  similar rest-frame colour  distributions suggesting similar
mass--to--light ratios, M/L.  Adopting $\rm \log M/L=+0.1$, consistent
with the colours of the galaxies in the sample (Bell \& de Jong 2001),
the  above mean optical  luminosities translate  to stellar  masses of
$\approx \rm  4 \times 10^{10}$ and $\rm  2 \times 10^{10}\,M_{\odot}$
for $\rm 24\,\mu m$-bright and faint sources respectively.

A plausible interpretation  of the results above is  that AGN activity
requires  a massive  BH and  some  gas to  fuel it  (Kauffmann et  al.
2003).  Massive galaxies often host large BHs, while star-formation is
an indication  of gas availability.  The  $24\,\mu m$-bright galaxies
with $\Delta  C > -0.15$  are typically luminous, $M_B  \la -20$\,mag,
and therefore  most likely  massive. The morphological  evidence above
also indicates  star-formation in a  large fraction of  these systems.
This  is contrary  to  $24\,\mu m$-faint  galaxies,  which are  either
quiescent and/or less massive  than $24\,\mu m$-bright sources. Galaxy
encounters, not  necessarily major mergers,  may still be  required in
the picture above to disturb the  gas to the galaxy centre and also to
trigger the formation of stars.

\subsection{The nature of AGN activity in transition zone and red galaxies}

The X-ray hardness ratio $\ga  -0.1$ estimated in sections 5.1 and 5.2
for  $\rm  24\,\mu  m$-bright   galaxies  with  $\Delta  C>-0.15$  and
post-starbursts can be attributed  to low-luminosity Compton thin AGN,
Compton  thick  sources (e.g.   Figure  \ref{fig_hr}), or  radiatively
inefficient  accretion  flows  (e.g.   Brand et  al.   2005).   Figure
\ref{fig_lxl24}  shows that  the  average $\rm  24\,\mu m$-bright  but
X-ray undetected  galaxy with $\Delta  C >-0.15$ has  X-ray luminosity
about a  factor 10  below the typical  X-ray detected AGN.   The X-ray
undetected galaxies lie in the same region of the parameter space with
moderately  obscured  low-luminosity AGN  and  Compton thick  sources,
suggesting that the observed stacked signal may be associated with one
or the other type of AGN activity,  or possibly with a mix of both. If
the mid-IR  luminosity of $24\,\mu  m$-bright galaxies with  $\Delta C
>-0.15$ is, on  average, dominated by star-formation, and  not by dust
heated  by  the  central  engine,  then the  moderately  obscured  low
luminosity  AGN scenario  is  most probable.   We  note however,  that
star-formation in the host galaxy is also likely to contribute or even
dominate the mid-IR emission of  both the Terashima et al.  comparison
sample and  the X-ray  sources in  the AEGIS survey.   In the  case of
Compton thick activity, the intrinsic AGN luminosity is expected to be
100--1000 times  brighter than the  observed one (e.g. Iwasawa  et al.
1997).  Deeply buried AGN  with luminosities approaching those of QSOs
($L_X \rm \approx 10^{44} \, erg \, s^{-1}$) among the red galaxies in
this study cannot be ruled.   Such objects are postulated to match the
spectrum of the  diffuse X-ray background (e.g.  Gilli  et al.  2007),
but,   currently,   few    have   been   securely   identified   (e.g.
Georgantopoulos \&  Georgakakis 2007;  Tozzi et al.   2006). Selection
methods  that  include mid-IR  wavelengths,  like  the  one used  here
($24\mu  m$-bright and $\Delta  C>-0.15$) ,  are capable  of detecting
this  population  (e.g.   Lacy  et  al.  2004;  Stern  et  al.   2005;
Martinez-Sansigre et al.  2006;  Donley et al.  2007).  X-ray stacking
analysis  has indeed  confirmed that  at least  some of  these methods
produce samples with hard mean  X-ray spectra that are consistent with
heavily  obscured AGN (e.g.   Daddi et  al. 2007;  Fiore et  al. 2007;
Georgantopoulos et al. 2007).

Brand et al.  (2005) also reported a hard stacked X-ray signal for red
galaxies in the range $0.3 < z < 0.9$ selected in the XBootes field of
the NOAO Deep Wide-Field Survey  (Kenter et al.  2005).  These authors
suggested that  the detected signal  can be interpreted  as unabsorbed
emission  from a  radiatively  inefficient accretion  (such as  ADAFs;
Narayan \& Yi  1994, 1995).  We explore this  possibility by comparing
the  mean Eddington ratio  of $24\mu  m$-bright galaxies  with $\Delta
C>-0.15$  with that of  M\,87, the  best studied  example of  a system
undergoing radiatively inefficient accretion. The central supermassive
BH  of  this  galaxy  has  been  shown  to  have  very  low  radiative
efficiency, corresponding  to an Eddington ratio  of $\approx 10^{-7}$
(Di  Matteo et  al.  2003).   The mean  optical absolute  magnitude of
$24\mu  m$-bright  galaxies  with  $\Delta C>-0.15$  is  $M_B  \approx
-20.8$\,mag,  which  corresponds  to  a  mean  stellar  mass  of  $\rm
4\times10^{10}\,M_{\odot}$  for a mass--to--light  ratio of  $\rm \log
M/L=+0.1$, consistent  with the average  colours of the  galaxies with
$\Delta C>-0.15$  (Bell \& de Jong  2001).  Assuming that  the bulk of
the stellar  mass is  associated with the  galaxy bulge and  using the
relation between BH mass  and bulge dynamical mass of H\"{a}ring \&
Rix  (2004),  we  estimate a  BH  mass  of  $\rm  6 \times  10^{7}  \,
M_{\odot}$.  This is  an upper limit as the  estimated stellar mass of
the $24\mu m$-bright population may have a large contribution from the
disk of these galaxies, which  has been ignored in this exercise.  The
mean  2-10\,keV   X-ray  luminosity  of  this   population  (Table  2)
corresponds to a bolometric  luminosity of $L_{bol}=3.5 \times 10^{42}
\rm \, erg \, s^{-1}$,  adopting the bolometric correction of Elvis et
al.   (1994).  This  is a  lower limit  in the  case of  Compton thick
activity.  The BH mass and  the $L_{bol}$ estimates above translate to
a  {\it lower}  limit  for  the Eddington  ratio  of $\approx  5\times
10^{-4}$, more than 3\,dex  higher than the corresponding quantity for
M\,87.  If  the hard  X-ray signal of  $24\mu m$-bright  galaxies with
$\Delta  C>-0.15$   is  associated  with   a  radiatively  inefficient
accretion mode, they have to be significantly less extreme than M\,87.
For comparison, X-ray {\it detected} AGN at $z\approx1$ have Eddington
ratios in  the range $\rm  10^{-4}-10^{-1}$ (e.g.  Bundy et  al. 2007;
Babic et al.  2007). The mean Eddington ratio  of the $24\mu m$-bright
galaxies with  $\Delta C>-0.15$  lies at the  low-end of  the interval
above,  suggesting that they  are dominated  by low  luminosity and/or
Compton Thick AGN.

\section{Discussion}

\subsection{AGN and transition zone galaxies}

In the  nearby Universe, $z  \la 0.1$ there  has been evidence  for an
association   between    {\it   optically}   selected    AGN,   recent
star-formation and  galaxies in the  transition zone from the  blue to
the red cloud.  Kauffmann et al. (2003) found that luminous AGN in the
SDSS  are  hosted by  early-type  bulge-dominated  galaxies that  have
stopped forming stars,  but only in the {\it  recent} past.  Martin et
al.  (2007)  extended this  study by constructing  the near-UV/optical
CMD of  galaxies at $z<0.1$ using  data from the  GALEX Medium Imaging
Survey  and the  SDSS.   The  density of  Seyfert-2s  in that  sample,
identified  by diagnostic emission  line ratios,  peaks in  the valley
between  the two  clouds. Our  work  extends these  results to  higher
redshift, $z \approx 0.7$, by  finding evidence for obscured AGN among
galaxies  in the vicinity  of the  valley of  the CMD  and in  the red
cloud.  Similar  results also  apply to the  X-ray {\it  detected} AGN
population.   X-ray sources  with $\Delta  C >-0.15$  have  an average
hardness  ratio of  $-0.09\pm0.07$,  while for  those  with $\Delta  C
<-0.15$ the mean hardness ratio is $-0.31\pm0.08$.

The incidence of  AGN among transition zone galaxies  suggests that BH
accretion may  play a  role in the  evolution of these  systems.  This
link has been highlighted recently  by Hopkins et al.  (2007) who show
that  the buildup  rate of  the  red cloud  mass function  is in  good
agreement with predictions based on the QSO luminosity function.  This
calculation assumes  either a fixed Eddington accretion  rate for QSOs
and the  local BH-host  mass empirical relation  or a  more physically
motivated model  for the QSO  evolution based on simulations  where BH
accretion  is  triggered  by   gas-rich  galaxy  mergers  (Hopkins  et
al. 2006a,  b).  Hopkins  et al. (2007)  also found that  the observed
rate  of galaxy  mergers and  their  distribution in  stellar mass  at
different redshifts are broadly consistent with the growth rate of the
red cloud  mass function.  The evidence above  suggests an association
between AGN activity, the transition of galaxies to the red cloud, and
mergers.   These catastrophic  events  can lead  to the  morphological
transformation of galaxies and at the same time offer an efficient way
for channeling  gas to the  nuclear regions of the  galaxy, triggering
the growth of the central BH.

Although our analysis broadly  supports an association between AGN and
galaxies on the move to the  red cloud, the evidence for a causal link
between the  two is less  clear. The X-ray  spectra of AGN in  the red
cloud (both  detections and stacking results) imply  at least moderate
amounts of obscuration  in a large fraction of  these systems. This is
counter-intuitive to a picture where AGN-driven outflows blow away the
gas and dust clouds from  the nuclear galaxy regions and suggests that
some gas  and dust clouds either remain  at or relax to  the centre of
the galaxy after the quenching  of the star-formation to form a torus.
Additionally,  major mergers  are not  supported by  the  data.  Minor
mergers or tidal disruptions may still play a role in the evolution of
$24\mu m$-bright  galaxies with $\Delta C >-0.15$.   Recent studies on
the evolution of  the star-formation since $z \approx  1$ also suggest
that  a decline  in the  major mergers  is not  the  dominant physical
mechanism driving the evolution.  Gas exhaustion in disk galaxies that
form stars  in a quiescent mode  may be more important  (e.g.  Bell et
al.  2005; Wolf  et al. 2005; Melbourne, Koo \&  Le Floch 2005; Noeske
et al. 2007a, b).  The  gas depletion scenario is also consistent with
the obscured AGN activity observed  in galaxies in the transition zone
and in the red cloud.

\subsection{AGN in post-starbursts}

In the local Universe, $z<0.1$,  an increasing body of evidence points
to a  link between black-hole  accretion and post-starbursts.   Yan et
al.  (2006)  performed a systematic search for  post-starbursts in the
SDSS using methods similar to  those described here.  A large fraction
of  the sources in  that sample  ($>70$ per  cent) show  weak emission
lines  consistent  with   low-luminosity  and/or  obscured  AGN,  e.g.
LINERs, Seyfert-2s, transition objects.  In a complementary study Goto
(2006) selected SDSS  galaxies with deep $\rm  H\delta$ absorption and
emission line signatures typical of AGN. These sources represent about
4 per cent of the AGN in  a volume limited sample, much higher than in
the overall SDSS population (0.2 per cent).

Our results  extend the association between AGN  and post-stabursts to
high  redshift,  $z \approx  0.8$.   Firstly,  there  is an  increased
fraction  of  X-ray  detected  AGN  among  the  post-starburst  galaxy
population.  Secondly, about  20 per cent of the  red-cloud AGN in the
redshift interval  $0.7 \la z  \la 0.9$, where post-starbursts  can be
identified, are hosted  by such galaxies.  Most of  the X-ray detected
post-starbursts  are  obscured,  with  an average  hardness  ratio  of
$\approx -0.09$. Thirdly, stacking  the X-ray photons at the positions
of  post-starbursts  that  are  not  individually  detected  at  X-ray
wavelengths, reveals a hard  mean X-ray spectrum and suggests obscured
AGN activity in the bulk of this population.

The  low redshift,  post-starburst sample  of  Yan et  al.  (2006)  is
comparable in terms of selection  to that presented here.  Most of the
emission line sources in that sample belong to the LINER class.  These
low-luminosity AGN often  show hard X-ray spectra and,  in many cases,
are  associated  with  moderate  column densities,  $\rm  N_H  \approx
10^{22} - 10^{23} \, cm^{-2}$, or even Compton thick systems where the
X-ray emission is dominated  by reflected radiation (e.g. Terashima et
al. 2002; Gonz\'alez-Mart\'in et al.   2006).  In the latter case, the
intrinsic  AGN power  is likely  to be  significantly higher  than the
observed one.  The evidence above suggests a similar nature for the $z
\approx 0.8$ post-starbursts for which our analysis shows a hard X-ray
spectrum in the mean. While the incidence of AGN in post-starbursts is
consistent with models where BH accretion is responsible for quenching
the star-formation  in galaxies, the fact  than many of  these AGN are
obscured means that  gas and dust clouds remain  in the nuclear galaxy
regions, possibly in a form of  a torus, after the AGN-driven blow-out
phase.

\subsection{Obscured AGN in the red  cloud} 

A striking  result from this  study is that  the red cloud  includes a
large  population of  obscured AGN,  both  above and  below the  X-ray
detection threshold  of the AEGIS {\it Chandra}  survey.  This extends
recent studies,  which have shown that  a large fraction  of the X-ray
{\it detected} AGN at $z\approx 1$, and certainly, the majority of the
obscured  ones,  are  hosted   by  red  galaxies  (e.g.   Georgakakis,
Georgantopoulos  \&  Akylas 2006;  Nandra  et  al.   2007; Rovilos  \&
Georgantopoulos 2007).

The incidence of  a large population of obscured AGN  in the red cloud
suggests   that  the   BH  accretion   outlives  the   termination  of
star-formation.   This is also  supported by  the X-ray  properties of
post-starburst  galaxies in our  study.  It  is not  clear why  such a
large  fraction of red  cloud galaxies  and post-starbursts  show high
column densities.   If these sources  represent systems after  the AGN
driven termination of star-formation then one would expect the central
engine to  be unobscured. The  X-ray obscuration in these  systems may
therefore represent  cold gas that either  has not been  blown out, or
has not been processed and which  relaxes to the galaxy centre to form
a torus surrounding the central  engine (e.g.  Hopkins et al.  2006a).
Alternatively, gas  exhaustion and not necessarily  outflows driven by
accreting BHs, may  be the mechanism behind the  evolution of both AGN
and galaxies. In  this picture the AGN may remain  obscured as long as
there are dust  and gas clouds to feed the central  BH.  Also, some of
the red cloud $\rm 24\mu m$ bright sources may be dusty systems (hence
the red colour) observed before the quenching of the star-formation.

The large fraction of obscured AGN in the red cloud is contrary to the
distribution  of {\it  optically} selected,  low redshift  AGN  in the
GALEX near-UV/optical CMD (Martin et  al. 2007). The number density of
Seyfert-2s in that study peaks in the valley.  This discrepancy may be
related to  selection effects. For  example, dilution of  the emission
lines by  the host galaxy stellar population  makes the identification
of the  AGN optical spectral  signatures difficult (e.g.   Comastri et
al.   2002; Severgnini  et al.   2003; Georgantopoulos  \& Georgakakis
2005).  Also, contamination of the GALEX near-UV band by scattered AGN
light  will move  red galaxies  into the  valley.   Alternatively, the
GALEX near-UV bands may be  more sensitive to low level star-formation
compared to optical wavebands. In any case, this potential discrepancy
highlights  the need  to  study in  more  detail the  overlap and  the
differences between  optical and X-ray AGN selection  methods in terms
of host galaxy properties.

\section{Conclusions}

We  explore  the  role  of  AGN in  establishing  the  bimodal  colour
distribution  of  galaxies by  quenching  the  star-formation of  blue
star-forming  systems  causing their  transformation  to red,  evolved
galaxies.  The main conclusions from this study are summarised below

  1. There  is  evidence  for  AGN  activity  among  galaxies  in  the
transition zone between  the red and blue clouds of  the CMD.  A large
fraction of these  accreting BHs are obscured, suggesting  that if AGN
outflows  are related  to the  colour transformation  of  galaxies, at
least some nuclear gas and dust  gas clouds are either not affected or
can efficiently reform after the truncation of the star-formation.

2. Morphological analysis suggests that  major mergers do not dominate
the  evolution of  this population.   Minor interactions  however, may
play a role.

3. We find an association  between BH accretion and post-starbursts at
$z \approx  0.8$, in agreement with  studies on the  properties of AGN
hosts at low redshift, $z \approx 0.1$.

4.  AGN  activity outlives  the termination  of the  star-formation. A
large fraction of  active BHs are present in  red cloud galaxies. This
is in contrast  to optically selected AGN, which  lie predominantly in
the valley between the two clouds.

\section{Acknowledgements}
We thank the referee, Rachel Somerville, for providing constructive
comments and suggestions that significantly improved this paper.
This  work  has  been   supported  by  funding  from  the  Marie-Curie
Fellowship  grant MEIF-CT-2005-025108  (AG), STFC  (ESL)  and Chandra
grant  GO5-6141A (DCK).  JML  acknowledges support  from the  NOAO Leo
Goldberg   Fellowship,   the   NASA   grants   HST-GO-10314.13-A   and
HST-AR-10675-01-A from the Space Telescope Science Institute, which is
operated by the AURA, Inc., under NASA contract NAS5-26555, NASA grant
NAG5-11513  to P.   Madau. JAN  is  supported by  NASA through  Hubble
Fellowship  grant  HF-011065.01-A,  awarded  by  the  Space  Telescope
Science   Institute,  which   is  operated   by  the   Association  of
Universities for Research in Astronomy, Inc., for NASA, under contract
NAS 5-26555.

Support for GO  program 10134 was provided by  NASA through NASA grant
HST-G0-10134.18-A from the Space Telescope Science Institute, which is
operated by the Association of Universities for Research in Astronomy,
Inc., under NASA contract NAS 5-26555.

The  authors wish to  recognise and  acknowledge the  very significant
cultural role  and reverence that the  summit of Mauna  Kea has always
had within the indigenous Hawaiian community. We are most fortunate to
have the opportunity to  conduct observations from this mountain. This
work  is based in  part on  observations made  with the  {\it Spitzer}
Space Telescope,  which is operated by the  Jet Propulsion Laboratory,
California Institute of Technology under a contract with NASA. Support
for  this  work  was provided  by  NASA  through  an award  issued  by
JPL/Caltech.


\begin{thebibliography}{}

	
\bibitem{} Abraham R. G., et al., 1996, ApJS, 107, 1


\bibitem{} Alexander D. M., et al., 2005, Nature, 434, 738

\bibitem{} Alexander D. M., et al., 2003, AJ, 126, 539

\bibitem{} Babic A., Miller L., Jarvis M. J., Turner T. J., Alexander
  D. M., Croom S. M., 2007, A\&A, 474, 755

\bibitem{} Baldry I. K., Glazebrook K., Brinkmann J., Ivezic Z.,
  Lupton R. H., Nichol R. C., Szalay A. S., 2004, ApJ, 600, 681 

\bibitem{} Barger A. J., et al., 2005, AJ, 129, 578

\bibitem{} Barmby P., et al., 2006, ApJ, 642, 126


\bibitem{} Bell E., Phleps S., Somerville R. S., Wolf C., Borch A.,
  Meisenheimer K., 2006a, ApJ, 652, 270

\bibitem{} Bell E., et al., 2006b, ApJ, 640, 241

\bibitem{} Bell E., et al., 2005, ApJ, 625, 23


\bibitem{} Bell E., et al., 2004, ApJ, 608, 752

\bibitem{} Blanton M. R., 2006, ApJ, 648, 268

\bibitem{} Brand K., et al., ApJ, 2005, 626, 723

\bibitem{} Brusa M., et al., 2007, ApJS, 172, 353

\bibitem{} Bundy K., et al., 2007, ApJ, submitted, arXiv0710.2105

\bibitem{} Cattaneo A., et al., 2007, MNRAS, 377, 63

\bibitem{} Cid Fernandes R., et al., 2001, ApJ, 558, 81

\bibitem{} Conselice C. J., Bershady M. A., Jangren A., 2000, ApJ, 529,
  886

\bibitem{} Comastri A., et al., 2002, ApJ, 571, 771

\bibitem{} Croton D. J, et al.,  2006, MNRAS, 365, 11

\bibitem{} Daddi E., et al., 2007, ApJ, submitted, arXiv:0705.2832

\bibitem{} Dale D. A., et al., 2007, ApJ, 655, 863

\bibitem{} Davis M., et al., 2007, ApJ, 660, 1L

\bibitem{} Dekel A., Birnboim Y., 2006, MNRAS, 368, 2


\bibitem{} Di Matteo T., Allen  S. W., Fabia A. C. Fabian, Wilson  A. S., Young
A. J., ApJ, 2003, 582, 133. 


\bibitem{} Donley J. L., et al., 2005, ApJ, 634, 169


\bibitem{} Elvis M.,  1994, ApJS, 95, 1


\bibitem{} Faber et al. 2003, SPIE, 4841, 1657

\bibitem{}  Fabian A. C., 1999, MNRAS, 308L, 39


\bibitem{} Ferrarese L. \& Merritt D., 2000, ApJ, 539, L9

\bibitem{} Franceschini A., et al., 2003, MNRAS, 343, 1181

\bibitem{} Fiore F., et al., 2007, ApJ, submitted, arXiv:0705.2864

\bibitem{} Gebhardt K., et al., 2000, ApJ, 539, L13

\bibitem{} Genzel R., et al., 1998, ApJ, 498, 579

\bibitem{} Georgakakis A., Georgantopoulos I. \& Akylas A., 2006,
  MNRAS, 366 ,171

\bibitem{} Georgakakis A., Hopkins A. M., Afonso J., Sullivan M., Mobasher
B., Cram L. E., 2004, MNRAS, 354, 127

\bibitem{} Georgantopoulos I., et al., 2007, MNRAS, submitted

\bibitem{} Georgantopoulos I. \& Georgakakis A., 2007, A\&A, 466, 823

\bibitem{} Georgantopoulos I. \& Georgakakis A., 2005, MNRAS, 358, 131

\bibitem{} Gerke B. F., et al.,  2007, MNRAS, 376, 1425

\bibitem{} Gilli R., Comastri A., Hasinger G., 2007, A\&A, 463, 79

\bibitem{}  Gonz\'alez-Mart\'in O., Masegosa J., M\'arquez I., Guerrero M. A.,
 Dultzin-Hacyan D.,  2006, A\&A, 460, 45

\bibitem{} Goto T., 2006, MNRAS, 369, 1765

\bibitem{} Grogin N. A., 2005, ApJ, 627, 97L

\bibitem{} Grogin N. A., et al., 2003, ApJ, 595, 685

\bibitem{} H\"aring N. \& Rix H. W., 2004, ApJ, 604L, 89 

\bibitem{} Hasinger G., Miyaji T., Schmidt M., 2005, A\&A, 441, 417


\bibitem{} Hopkins P. F.,  et al., 2005, ApJ,  630, 705

\bibitem{} Hopkins P. F., Hernquist L., Cox T. J., Di Matteo T., Robertson B.,
S. Volker,  2006a, ApJS, 163, 1

\bibitem{} Hopkins P. F., Somerville R. S., Hernquist L., Cox T. J., Robertson
B., Li Y.,  2006b, ApJ, 652, 864

\bibitem{} Hopkins P. F., Bundy K., Hernquist L., Ellis R. S.,  
2007, ApJ, 659, 976

\bibitem{} Iwasawa K., Fabian A. C., Matt G., 1997, MNRAS, 289, 443

\bibitem{} Kauffmann G., et al., 2004, MNRAS, 353, 713

\bibitem{}  Kauffmann G., et al., 2003, MNRAS, 346, 1055

 \bibitem{}  Kenter A., et al., 2005, ApJS, 161, 9

\bibitem{} King A.,  2003, ApJ, 596, 27L

\bibitem{} Cirasuolo M., et al., 2007, MNRAS, submitted, arXiv:astro-ph/0609287

\bibitem{}  Lacy  M.,  et  al.,  2004, ApJS,  154,  166


\bibitem{} Laird E. S., Nandra K., Adelberger K. L., Steidel C. C., Reddy N. A.,
2005, MNRAS, 359, 47

\bibitem{} Lotz J. M., et al., 2007, ApJ, submitted, arXiv:astro-ph/0602088 
	
\bibitem{} Lotz J. M., Primack J., Madau P.,  2004, AJ, 128, 163

\bibitem{} Magorrian J., et al., 1998, AJ, 115, 2285

\bibitem{} Martin C. D., et al., 2007, ApJS, in press, astro-ph: 0703281

\bibitem{} Martinez-Sansigre  A., et al.,  2006, MNRAS,  370, 1479

\bibitem{} Matt G., et al., 1999, A\&A, 341, 39L


\bibitem{} Melbourne J., Koo D. C., Le Floch E., 2005, ApJ, 632L, 65

\bibitem{} Nandra K., et al., 2007, ApJ, 660, 11L

\bibitem{} Nandra K., et al., 2005, MNRAS, 356, 568

	
\bibitem{}  Nandra K., Mushotzky R. F., Arnaud K., Steidel C. C., Adelberger
K. L.,  Gardner J. P., Teplitz H. I., Windhorst R. A., 2002, ApJ, 576,
625
	
\bibitem{} Nandra K. \& Pounds K. A.,  1994, MNRAS, 268, 405

\bibitem{} Narayan R., Igumenshchev I. V., \& Abramowicz M. A. 2000, ApJ, 539, 798 

\bibitem{}Narayan R., \& Yi I., 1994, ApJ, 428, L13

\bibitem{}  Noeske K. G. et al., 2007a ,ApJ, 660L, 47

\bibitem{}  Noeske K. G. et al., 2007b ,ApJ, 660L, 43

\bibitem{} Okamoto T., Nemmen R. S., Bower R. G., 2007, MNRAS, submitted, arXiv:0704.1218

\bibitem{} Pierce C. M., et al., 2007, ApJ, 660, 19L

\bibitem{} Ptak A., Heckman T., Levenson N. A., Weaver K., Strickland D., 2003,
ApJ, 592, 782

\bibitem{} Quintero A. D., et al. 2004, ApJ, 602, 190

\bibitem{} Ranalli P., Comastri A., Setti G., 2003, A\&A, 399, 39  

\bibitem{} Rovilos E., Georgakakis A., Georgantopoulos I., Afonso J.,
  Koekemoer A. M., Mobasher B., Goudis C.,  2007, A\&A, 466, 119


\bibitem{} Rovilos E. \& Georgantopoulos I., 2007, A\&A, 475, 115

\bibitem{} Severgnini P., et al., 2003, A\&A, 406, 483

\bibitem{} Silk J., \& Rees M. J.	1998, A\&A, 331, 1L

\bibitem{} Stern D.,  et al., 2005, ApJ, 631, 163

\bibitem{} Strateva I., 2001, AJ, 122, 1861


\bibitem{} Terashima Y., Iyomoto N., Ho L. C., Ptak
  A. F.,   2002, ApJS, 139, 1

\bibitem{} Tozzi P., et al., 2006, A\&A, 451, 457



\bibitem{} Weiner B. J.,  2005, ApJ, 620, 595
\bibitem{} Willmer C. N. A., et al.,  2006, ApJ, 647, 853

\bibitem{} Wolf C., et al., 2005, ApJ, 630, 771

\bibitem{} Yan R. et al., 2006, ApJ, 648, 281

 \bibitem{} Yan L. et al., 2004, ApJS, 154, 60
\end{thebibliography}
\end{document}